\documentclass[sigconf]{acmart} 

\usepackage[notheorems]{eda} 
\usepackage[algo2e,linesnumbered,ruled,vlined]{algorithm2e}
\usepackage{multirow}
\usepackage{caption}
\usepackage{subcaption}
\setcitestyle{numbers,sort&compress}

\usepackage{xspace}
\newcommand{\ourmethod}{\textsc{Momo}\xspace} 
\newcommand{\ourmaintitle}{Graph Similarity Description: How Are These Graphs Similar?}
\newcommand{\ourtitle}{\ourmaintitle}
\newcommand{\oursummarizer}{\textsc{Beppo}\xspace} 
\newcommand{\ouralignment}{\textsc{Gigi}\xspace} 
\newcommand{\vog}{\textsc{VoG}\xspace} 
\newcommand{\maximalgreedy}{\textsc{MaximalGreedy}\xspace}

\newcommand{\oururl}{\url{http://eda.mmci.uni-saarland.de/prj/momo}}
\newcommand{\ourdoi}{\url{https://doi.org/10.5281/zenodo.4780912}}

\newcommand{\LN}{L_{\mathbb{N}}}

\newcommand{\type}{\text{type}}

\graphicspath{ {./figs/} }

\AtBeginDocument{%
  \providecommand\BibTeX{{%
    \normalfont B\kern-0.5em{\scshape i\kern-0.25em b}\kern-0.8em\TeX}}}

\setcopyright{rightsretained}

\acmSubmissionID{rst1537}

\begin{document}
\fancyhead{}

\newtheorem{problem}{Problem}

\title[\ourtitle]{\ourmaintitle}

\author{Corinna Coupette}
\affiliation{%
  \institution{Max Planck Institute for Informatics}
  \city{Saarbr\"ucken}
  \country{Germany}}
\email{coupette@mpi-inf.mpg.de}

\author{Jilles Vreeken}
\affiliation{%
  \institution{CISPA Helmholtz Center for Information Security}
  \city{Saarbr\"ucken}
  \country{Germany}}
\email{jv@cispa.de}

\renewcommand{\shortauthors}{Coupette and Vreeken}

\begin{abstract}

How do social networks differ across platforms? 
How do information networks change over time?
Answering questions like these requires us to compare two or more graphs. 
This task is commonly treated as a \emph{measurement} problem, 
but numerical answers give limited insight.
Here, we argue that if the goal is to gain understanding, we should treat graph similarity assessment as a \emph{description} problem instead.
We formalize this problem as a model selection task using the Minimum Description Length principle, 
capturing the similarity of the input graphs in a \emph{common model} and the differences between them in \emph{transformations} to \emph{individual models}.
To discover good models, we propose \ourmethod, which breaks the problem into two parts and introduces efficient algorithms for each.
Through an extensive set of experiments on a wide range of synthetic and real-world graphs, we confirm that \ourmethod works well in practice.
\end{abstract}

\copyrightyear{2021} 
\acmYear{2021} 
\acmConference[KDD '21]{Proceedings of the 27th ACM SIGKDD Conference on Knowledge Discovery and Data Mining}{August 14--18, 2021}{Virtual Event, Singapore}
\acmBooktitle{Proceedings of the 27th ACM SIGKDD Conference on Knowledge Discovery and Data Mining (KDD '21), August 14--18, 2021, Virtual Event, Singapore}
\acmDOI{10.1145/3447548.3467257}
\acmISBN{978-1-4503-8332-5/21/08}




\maketitle

\section{Introduction}
\label{sec:intro}

Comparing two or more graphs is important in many applications. 
In biology, we might, for example, want to compare the protein interaction networks of different human tissues so as to discover common and specialized mechanisms, while in the social sciences, comparing collaboration networks over time or across fields could reveal knowledge dynamics. 
The task of comparing graphs is called \emph{graph similarity assessment}. 
It is commonly treated as a \emph{measurement} problem, i.e., a question to which a numerical answer suffices (e.g., $0.42$). 
While such an answer may be useful in certain downstream tasks like classification or clustering, it provides limited insight and is thus generally dissatisfying to a domain expert. 

In this paper, we argue that if the goal is to gain understanding, we should not ask ``how \emph{similar} are these graphs?''
but rather ``\emph{how} are these graphs similar?''.
That is, 
we propose to treat graph similarity assessment as a \emph{description} problem, demanding an answer that, in easily understandable terms, characterizes what is similar and what is different between the input graphs at hand. 
We formalize the problem in information-theoretic terms using the Minimum Description Length (MDL) principle, 
by which we are after the shortest lossless description of the input graphs using common and specific structures (e.g., stars, cliques, bicliques, and starcliques) as well as shared nodes and edges between these structures. 
Since we can measure how many bits we gain by compressing the graphs jointly, rather than individually, our formalization also allows for an easily interpretable quantification of differences.

\begin{figure}
	\centering
	\includegraphics[width=0.95\linewidth]{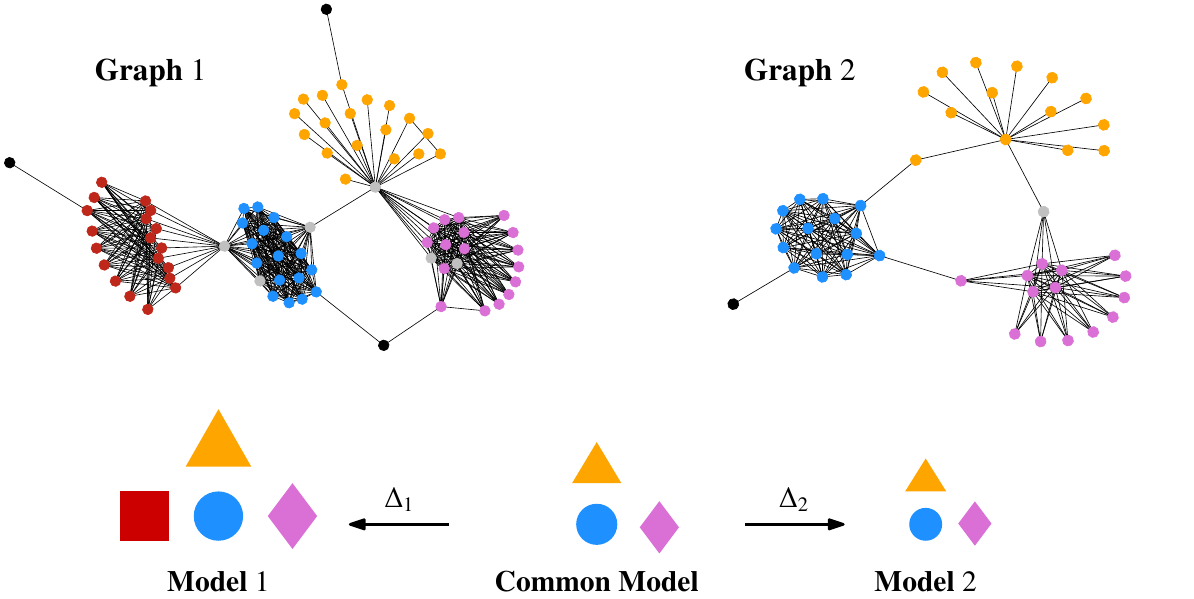}
	\vspace*{-8pt}\caption{A common model captures the structure shared between the individual models of the input graphs.}
	\label{fig:common-model}
\end{figure}

As an example of graph similarity description, consider Figure~\ref{fig:common-model}, which depicts two toy graphs and the result returned by our method. 
Even though the graphs are of different sizes, and no node alignment is given, our method discovers that both graphs contain a star (orange triangle) that is connected to a clique (blue circle) and a starclique (pink diamond). 
We further see that the left graph is different in that it additionally contains a biclique (red square), and that the structures in the left graph all contain more nodes than their counterparts in the right graph (larger shapes). 

\addtolength{\belowcaptionskip}{-5pt}

\addtolength{\belowcaptionskip}{5pt}

When assessing the similarity between graphs in practice, we face a very large search space: 
There exist exponentially many sets of nodes, i.e., potential structures, exponentially many \emph{sets} of structures, 
and---unless a full node alignment is given---also exponentially many alignments between the graphs. 
As our score exhibits no structure that we could exploit to efficiently discover the optimal solution, we propose a framework, called \ourmethod (\emph{Mo}del of \emph{mo}dels), 
that breaks the problem into two parts and introduces efficient algorithms for each: 
\oursummarizer discovers interpretable summaries for the individual input graphs,
and \ouralignment uses them to unveil their shared and specific structures, 
from which we can also compute an informative similarity score. 
Through an extensive set of experiments on a wide range of synthetic and real-world graphs, we confirm that our algorithms perform well in practice: 
We discover summaries that are useful for domain experts, 
identify meaningful similarities between the protein interaction networks of different human tissues, 
and reveal distinct temporal dynamics in the collaboration networks of different scientific communities. 
Not unimportantly, in practice, our methods scale near-linearly in the number of edges.

The remainder of the paper is structured as follows. Section~\ref{sec:prelim} introduces our notation and gives a primer on MDL, and Sections~\ref{sec:theory}, \ref{sec:algo}, and \ref{sec:exps} present our main contributions. 
We cover related work in Section~\ref{sec:related}, and round up with discussion and conclusions in Section~\ref{sec:conclusion}.
All our data, code, and results are publicly available,\!\footnote{\oururl; \ourdoi}
and further information for reproducibility is given in Appendix~A.

\section{Preliminaries}
\label{sec:prelim}

We consider graphs $G_i=(V_i,E_i)$ with $n_i = |V_i|$ nodes, $m_i=|E_i|$ edges, and adjacency matrix $\mathbf{A}_i$, omitting the subscripts when clear from context.
An alignment $\mathcal{A}_{ij}$ between the graphs $G_i$ and $G_j$, denoted $G_i\parallel_{\mathcal{A}}G_j$, 
is a bijection from $V_i$ to $V_j$. 
To allow comparisons between graphs of different sizes or graphs for which no node alignment is known,
we allow this bijection to be \emph{partial} or \emph{empty}, i.e., 
there can be nodes in $V_i$ ($V_j$) that have no image (preimage) in $V_j$ ($V_i$) under $\mathcal{A}_{ij}$.
We assume that our input graphs are \emph{simple}, i.e., undirected, unweighted, without loops or parallel edges, 
and that only \emph{two} input graphs are given, but our framework generalizes to comparisons between \emph{more than two} \emph{general} graphs.

We build on the notion of \emph{Kolmogorov complexity}.
The Kolmogorov complexity of an object $x$, $K(x)$, is the length in bits of the shortest program computing $x$ on a universal Turing machine, and the \emph{conditional} Kolmogorov complexity of $x$ given $y$, $K(x\mid y)$, 
is the length of such a program with $y$ as auxiliary input \cite{vitanyi:93:book}.
The \emph{Information Distance} between $x$ and $y$ is (up to an additive logarithmic term) the length of the shortest program transforming $x$ into $y$ and $y$ into $x$, i.e.,
$ID(x,y) = \max\{K(x\mid y),K(y\mid x)\}$ \cite{li:04:similarity}. 
Dividing by $\max\{K(x),K(y)\}$, we obtain the \emph{Normalized Information Distance}.

The Kolmogorov complexity is not computable, and hence, neither is the Normalized Information Distance.
To \emph{describe} and \emph{measure} the similarity between graphs in practice, we thus instantiate Kolmogorov complexity through the Minimum Description Length (MDL) principle \cite{grunwald:07:book}. 
MDL is a practical version of Kolmogorov complexity embracing the slogan \emph{Induction by Compression}. 
Given a model class $\mathfrak{M}$ for some data $\mathcal{D}$, 
the best model $M\in \mathfrak{M}$ minimizes
$L(M)+L(\mathcal{D}\mid M)$,
where $L(M)$ is the description length of $M$, $L(\mathcal{D}\mid M)$ is the description length of the data when encoded using $M$, 
and both are measured in bits under our encoding.
This is called \emph{crude} MDL, 
and it contrasts with \emph{refined} MDL, which encodes the model and the data together \cite{grunwald:07:book}. 
We opt for crude MDL not only because it is computable 
but also because we are particularly interested in the model: 
the structures shared by our input graphs, 
and the transformations necessary to derive the individual graphs from them. 
Finally, we require \emph{lossless} descriptions to ensure fair comparisons between competing models.

All logarithms are to base $2$, and we define $\log 0 = 0$. 
We use $\lfloor \cdot \rceil$ for rounding to the closest integer, and summarize our notation in Table~2 in the Appendix.

\section{Theory}
\label{sec:theory}

We now describe our first contribution, 
the MDL formulation of graph similarity assessment.
Our data is $\mathcal{D} = (G_1,G_2,\mathcal{A})$, 
where $G_1$ and $G_2$ are our input graphs,
and $\mathcal{A}$ is a (potentially partial or empty) node alignment between $G_1$ and $G_2$.

\subsection{Similarity Description, Informally}
\label{subsec:problem-informal}

Our primary goal is to \emph{describe} the similarity of our input graphs.
That is, we aim to find the key structures that are shared between these graphs and contrast them with the structures that are specific to the individual graphs. 
By \emph{structures}, we mean subgraphs whose connectivity follows distinct, interpretable patterns. 
Our \emph{structure vocabulary} $\Omega$ comprises four structure types: 
(approximate) \emph{cliques}, \emph{stars}, \emph{bicliques}, and \emph{starcliques}.
We choose these structure types because they are simple and widespread in real-world graphs from many different fields, but further structure types can easily be included, e.g., to tailor our method to a particular domain.

Intuitively, \emph{cliques} are subgraphs with relatively homogeneous connectivity whose density stands out against the background distribution (e.g., echo chambers in social networks).
\emph{Stars} are subgraphs in which one node, the \emph{hub}, is connected to all other nodes, the \emph{spokes}, and the spokes are hardly connected among themselves (e.g., influencers and their followers).
\emph{Bicliques} are subgraphs whose nodes can be partitioned into two sets, \emph{left} ($L$) and \emph{right} ($R$), such that 
$L$ and $R$ are densely connected, 
the nodes in $L$ are sparsely interconnected, 
and the nodes in $R$ are sparsely interconnected (e.g., predators and prey in food webs).
\emph{Starcliques} are bicliques whose left nodes are densely, rather than sparsely, interconnected---%
i.e., stars whose hub is a clique (e.g., core and periphery in infrastructure networks).
To describe real-world graphs accurately, we allow structures to overlap on nodes \emph{and} on edges.

\addtolength{\belowcaptionskip}{-5pt}

\begin{figure}
	\centering
	\includegraphics[width=0.95\columnwidth]{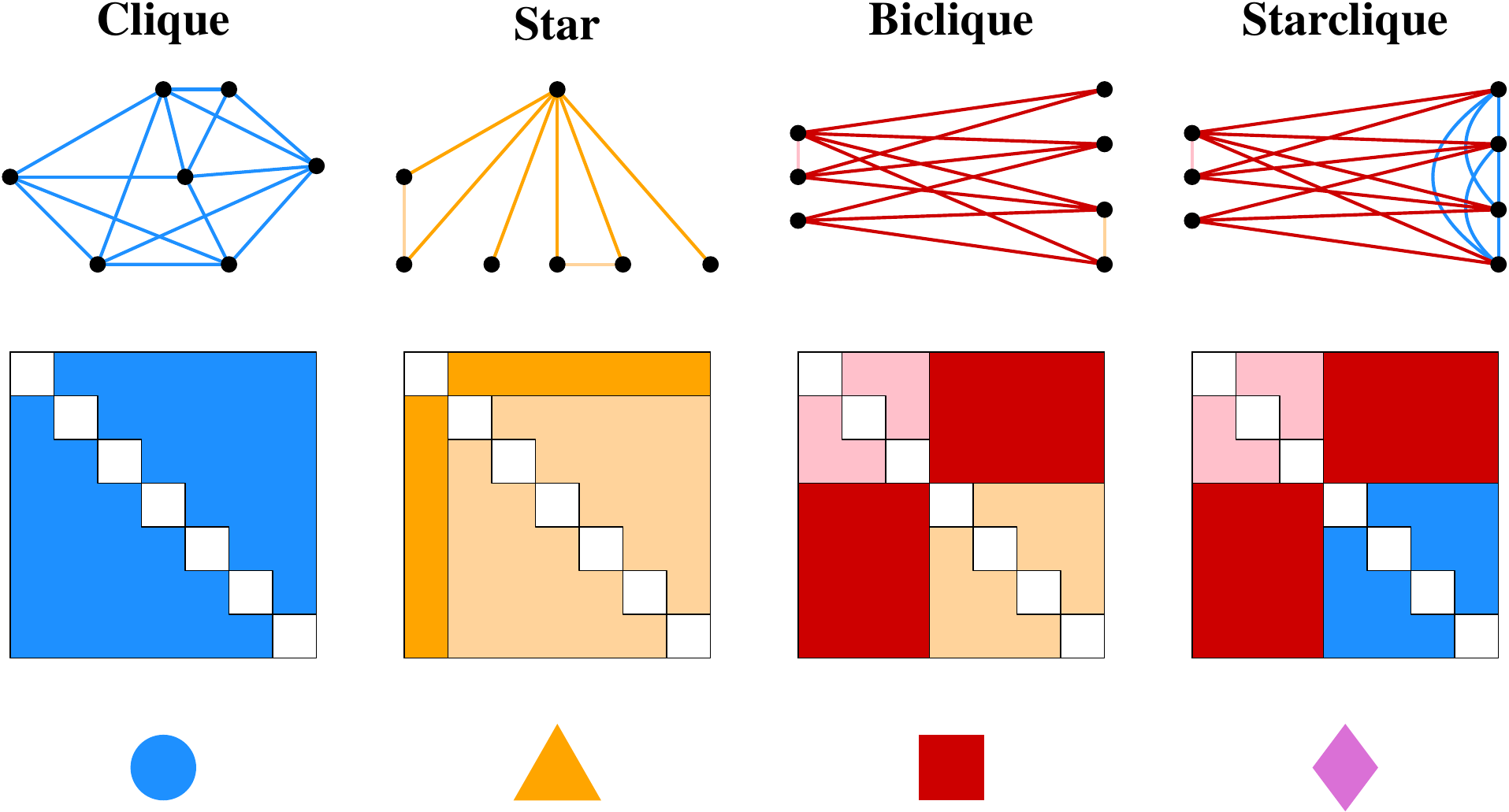}
	\vspace*{-8pt}\caption{%
		Graph structures are constraints over sets of (non-) edges. 
		They can be visualized as induced subgraphs (top), adjacency submatrices (middle), or shapes (bottom). 
		Each color in the adjacency submatrix is associated with a different constraint, where the white constraint enforces loop-freeness.
}\label{fig:structures}
\end{figure}

\addtolength{\belowcaptionskip}{5pt}

As depicted in Figure~\ref{fig:structures},
each structure imposes a set of constraints on the connectivity in the adjacency submatrix it identifies.
We think of the node set sizes of a structure as \emph{node fractions} (relative to a reference $n$) and of its connectivity constraints as \emph{edge densities} (relative to the maximum possible number of edges).

We represent the structures we find in $G_1$ and $G_2$ individually as lists $S_1$ and $S_2$ in their \emph{individual models} $M_1$ and $M_2$,
and the structures that are shared between $G_1$ and $G_2$ as a list $S_{12}$ in their \emph{common model} $M_{12}$.
To decide which structures to include in $S_{12}$, 
we construct a matching $\mathcal{M}\subseteq S_1\times S_2$, 
requiring that matched structures have the same type.
For each $(s_1,s_2)\in \mathcal{M}$, we include one structure $s$ of its type in $S_{12}$, writing $\varphi_1(s) = s_1$ and $\varphi_2(s) = s_2$ for the mappings from the shared structures to their counterparts.
The node fractions (edge densities) of $s$ are the averages of the node fractions (edge densities) in $s_1$ and $s_2$.
For example, if $s_1\in S_1$ is a clique with node fraction $0.1$ and edge density $0.9$, 
and $s_2\in S_2$ is a clique with node fraction $0.2$ and edge density $0.7$, 
$s\in S_{12}$ is a clique with node fraction $0.15$ and edge density $0.8$.

To link the common model to the individual models, we translate $M_{12}$ into $M_1$ and $M_2$ using \emph{transformations} $\Delta_1$ and $\Delta_2$, i.e., $\Delta_1(M_{12}) = M_1$ and $\Delta_2(M_{12}) = M_2$. 
Our \emph{transformation vocabulary} $\Sigma$ contains edit operations to (1) add unmatched structures contained in individual models, and (2) morph structures from $M_{12}$ into those from $M_1$ and $M_2$, 
i.e., reverse the averaging we perform when specifying the shared structures.
For example, if a clique $s\in S_{12}$ has node fraction $0.15$ and edge density $0.8$, 
and $\varphi_1(s) = s_1 \in S_1$ has node fraction $0.1$ and edge density $0.9$, 
we need to \emph{shrink} the node fraction and \emph{grow} the edge density of $s$ to match those in $s_1$. 

To discover our common model $M_{12}$, individual models $M_1$, $M_2$,
and transformations $\Delta_1$, $\Delta_2$, we leverage the MDL principle. 
We seek to optimize $L(M_{12}) + L(\Delta_1,\Delta_2) + L(G_1\parallel_{\mathcal{A}} G_2\mid M_{12},\Delta_1,\Delta_2)$.
For this purpose, we have to define several encodings.

\subsection{Similarity Description Encodings}
\label{subsec:define-encodings}
 
We need to describe how we encode 
(1) the graphs $G_1$, $G_2$ under their individual models $M_1$, $M_2$, 
(2) the models $M_1$, $M_2$, 
(3) the common model $M_{12}$, 
and (4) the transformations $\Delta_1$, $\Delta_2$ in bits.

\subsubsection*{Encoding a Graph Under an Individual Model}

Given a model $M$ of a graph $G$, rather than using an ad-hoc encoding of the graph under the model (as is common practice),
we seek to encode $G$ \emph{optimally}, leveraging the knowledge contained in $M$.
As depicted in Figure~\ref{fig:structures}, 
this knowledge primarily comes as constraints on the total number of edges in the parts of the adjacency submatrix identified by the structures in $M$:
A clique imposes one constraint, a star imposes two constraints, and a biclique or starclique imposes three constraints.

The probability distribution over the adjacency matrix $\mathbf{A}$ of $G$ that represents the knowledge imparted by $M$ (which includes $n$, $m$, and loop-freeness) \emph{without any bias} is the distribution with the largest entropy among all distributions fulfilling the constraints imposed by $M$. 
Under this \emph{maximum entropy distribution},
$\Pr(a_{ij}  \mid M) = \frac{\exp(\sum_{\lambda\in\Lambda(i,j)} \lambda)}{1 + \exp (\sum_{\lambda\in\Lambda(i,j)} \lambda)}$,
where $\Lambda(i,j)$ is the set of Lagrange multipliers associated with the constraints covering $a_{ij}\in\mathbf{A}$
in the optimization problem finding the maximum entropy distribution for $\mathbf{A}$ given $M$.
The Shannon-optimal code based on this distribution minimizes the worst-case expected length of a message coming from the true distribution \cite{debie:11:dami}.
Hence, the length of $G$ given $M$ under an optimal encoding is
\begin{align*}
    L(G\mid M) = \sum_{a_{ij} \in A_1} -\log \Pr(a_{ij}  \mid M) + \sum_{a_{ij} \in A_0} - \log (1- \Pr(a_{ij} \mid M))\hphantom{\;},
\end{align*}
where $A_x = \{a_{ij} \in \mathbf{A} \mid (a_{ij} = x) \wedge (i < j)\}$ for $x\in \{0,1\}$.

\subsubsection*{Encoding an Individual Model}

To encode an individual model $M$ for a graph $G$, we communicate 
$n$, $m$, and $|S|$ using $L_{\mathbb{N}}$, the universal code for positive integers \cite{rissanen:83:integers}.
We then transmit the number of structures per type, 
and for each structure, in order, its type and its length. 
Thus, the length of an individual model $M$ for a graph $G$ is
\begin{align*}
	L(M) = \LN(n+1) + \LN(m+1) + \LN(|S|+1) + \log\binom{|S|+|\Omega|-1}{|\Omega|-1}\\
	+ \sum_{s\in S}\big(-\log \Pr(\type(s)\mid S) + L(s)\big)&\hphantom{\;}.
\end{align*}
Each structure is defined \emph{abstractly} by its constraints (cf. Figure~\ref{fig:structures}), 
and when we seek to find an MDL-optimal individual model, it is further identified by \emph{concrete} node IDs (typeset in grey).
Assuming that all structures contain a positive number of nodes,
the detailed encoding of our structures is as follows.

\paragraph*{Cliques}
To communicate a clique $s$, we transmit 
its number of nodes $n_s$, 
its number of edges $m_s$ or non-edges $\overline{m}_s$,
and the node IDs.
Therefore, with $m^*_s = n_s(n_s-1)/2$, the length of a clique is
\begin{align*}
	L(s) = \LN(n_s)
	+ 1 + \log\log\bigg\lfloor\frac{m^*_s}{2}\bigg\rfloor + \log(\min\{m_s,\overline{m}_s\})
	~\textcolor{black!60}{+ \log\binom{n}{n_s}}\hphantom{\;}.
\end{align*}

\paragraph*{Stars}
To communicate a star $s$, we transmit
its number of spokes $n_s-1$,
the number of edges between its spokes $x_s = m_s-n_s+1$,
the hub's ID, and the spokes' IDs.
Hence, with $x^*_s = (n_s-1)(n_s-2)/2$, the length of a star is
\begin{align*}
	L(s) = \LN(n_s-1) + \log\log x^*_s + \log x_s 
	\textcolor{black!60}{~+ \log n + \log\binom{n-1}{n_s-1}} \; .
\end{align*}

\paragraph*{Bicliques and Starcliques}
To communicate a biclique $s$, we transmit
(1) its number of nodes $n_s$,
(2) its number of left nodes $n_L$,
(3) its number of edges between left nodes $m_L$, 
(4) its number of edges between right nodes $m_R$, 
(5) its number of non-edges between left nodes and right nodes $m^*_A-m_A$ (where $m^*_A:=n_Ln_R$), and
(6) the IDs of its left nodes and its right nodes.
Thus, with $m^*_L = n_L(n_L-1)/2$ and $m^*_R = n_R(n_R-1)/2$, the length of a biclique is
\begin{align*}
	L(s) = \LN(n_s)
	+ \log n_s
	+ \log\log m^*_L + \log m_L
	+ \log\log m^*_R + \log m_R\\
	+ \log\log m^*_A + \log (m^*_A-m_A)
	~\textcolor{black!60}{+ \log\binom{n}{n_L}
	+ \log\binom{n - n_L}{n_s - n_L}}&\hphantom{\;}.
\end{align*}
To transmit a starclique $s$, we replace $m_L$ by $\overline{m}_L = m^*_L-m_L$.

\subsubsection*{Encoding a Common Model}

When communicating $M_{12}$, w.l.o.g., we assume that $n_1 \geq n_2$,
and we transmit the node fractions and edge densities of all shared structures with reference to $n_1$.
Since we explicitly want to handle unaligned graphs and graphs of different sizes, 
their common model does not include node IDs.
To encode $M_{12}$, we hence use the expression for individual models, 
with the node ID parts omitted, and the terms for $n$ and $m$ replaced by 
\begin{displaymath}
	\LN(n_1+1) 
	+ \LN(n_1-n_2+1)
	+ \LN(m_1+1)
	+ \LN(|m_1-m_2|+1) + 1\hphantom{\;}.
\end{displaymath}

\subsubsection*{Encoding Transformations}

The common model $M_{12}$ contains only structures that are shared between $G_1$ and $G_2$, and structures may be shared without being isomorphic. 
Consequently, $M_{12}$ is generally different from $M_1$ and $M_2$, even if we define all models without node IDs.
\emph{Transformations} link $M_{12}$ to $M_1$ and $M_2$ such that $\Delta_1(M_{12}) = M_1$ and $\Delta_2(M_{12}) = M_2$. 
That is, for $i\in\{1,2\}$, $\Delta_i$ morphs $M_{12}$ into $M_i$ by \emph{growing} or \emph{shrinking} the node fractions and edge densities of the structures in $S_{12}$ to match those in $S_i$ as well as \emph{adding} those structures from $S_i$ that have no counterpart in $S_{12}$.

To derive the necessary content for the transformations, we reason as follows.
The node fractions and edge densities of each structure $s\in S_{12}$ are the average of its representatives in $S_1$ and $S_2$, $\varphi_1(s)$ and $\varphi_2(s)$. 
Hence, for each structure in $s\in S_{12}$, we expect a structure of the same type in $S_1$ and $S_2$. 
For each node fraction $x$ in $s$, we expect the size of its counterpart in $\varphi_i(s)$ to be $\lfloor x\cdot n_i\rceil$, 
and for each edge density $y$ in $s$, we expect the number of edges in its counterpart in $\varphi_i(s)$ to be $\lfloor y \cdot m^*_y \rceil$, where $m^*_y$ is the maximum number of edges in the associated area of $\mathbf{A}_i$ (for $i\in\{1,2\}$).

The transformation $\Delta_i$ is the deviation of $M_i$ from our expectation based on $M_{12}$, 
and since the node fractions and edge densities of each structure in $S_{12}$ are the average of its representatives in $S_1$ and $S_2$, for the shared structures, we can infer $\Delta_2$ from $\Delta_1$.
Hence, to communicate $\Delta_1$ and $\Delta_2$, for each node fraction $x$ (edge density $y$) in each structure $s\in S_{12}$, we transmit the number of nodes (edges) we need to add or subtract from $\lfloor x\cdot n_1\rceil$ ($\lfloor y \cdot m^*_y \rceil$) to arrive at the size of its counterpart in $\varphi_1(s)$, along with the change direction (\emph{grow} or \emph{shrink}).
Finally, we transmit the structures in $\overline{S}_1 := S_1\setminus~\varphi_1(S_{12})$ and the structures in $\overline{S}_2 := S_2\setminus~\varphi_2(S_{12})$.

Therefore, if $L(\delta_1: \delta_1(s)=\varphi_1(s))$ is the description length of the transformation $\delta_1$ morphing $s$ into $\varphi_1(s)$, and
$T$ is the total number of change directions we need to transmit, 
the length of the transformations $\Delta_1$ and $\Delta_2$ is
\begin{align*}
	L(\Delta_1,\Delta_2) = \sum_{s\in S_{12}} L\big(\delta_1: \delta_1(s)=\varphi_1(s)\big)
	+ \log T + \sum_{i\in\{1,2\}}\LN(|\overline{S}_i|+1)&\\
	+ \sum_{i\in \{1,2\}}\Bigg(\log\binom{|\overline{S}_i|+|\Omega|-1}{|\Omega|-1}
	+ \sum_{s\in \overline{S}_i} \big(-\log\Pr(\type(s)\mid \overline{S}_i) + L(s)\big)\Bigg)&\hphantom{\;}.
\end{align*}

Although we have defined the individual models $M_1$ and $M_2$, the common model $M_{12}$, and the transformations $\Delta_1$ and $\Delta_2$ for similarity \emph{description}, we can also use them for similarity \emph{measurement}.

\subsection{Similarity Measurement}
\label{subsec:define-score}
For similarity measurement, our score should reflect the extent to which the structure of the input graphs can be captured by their common model. 
Since graphs have many permutation-invariant representations (unlike, e.g., strings), 
and computable instantiations of the Normalized Information Distance typically use opaque compressors, 
defining such a score is not straightforward.
To guarantee computability and interpretability, we thus instantiate the Normalized Information Distance using our \emph{models as compressors}.  

Let $G_1$ and $G_2$ be our input graphs with alignment $\mathcal{A}$ and individual models $M_1$ and $M_2$ (encoded without node IDs). 
Let $M_{12}$ be their best $\mathcal{A}$-respecting common model, 
and let $\Delta_1$ and $\Delta_2$ be transformations such that $\Delta_1(M_{12})=M_1$ and $\Delta_2(M_{12})=M_2$.
The \emph{Normalized Model Distance} (NMD) between $G_1$ and $G_2$ is
\begin{displaymath}
	\mathrm{NMD}(G_1,G_2) = \frac{L(M_{12})+L(\Delta_1,\Delta_2)-\min\{L(M_1),L(M_2)\}}{\max\{L(M_1),L(M_2)\}}\hphantom{\;}.
\end{displaymath}
The NMD is $0$ if $M_{12} = M_1 = M_2$ (with $\Delta_1=\Delta_2 = \emptyset$), and it is $1$ if $M_{12}=\emptyset$ (with $\Delta_1 = M_1$ and $\Delta_2 = M_2$).
It allows us to compare our method with other similarity \emph{measurement} methods even though our primary goal is similarity \emph{description}, which we formalize next.

\subsection{Similarity Description, Formally}
\label{subsec:problem-formal}

We are now ready to formally state our problem.

\vspace*{0.7em}\noindent\textsc{\bfseries The Graph Similarity Description Problem}.
\emph{Given graphs $G_1$, $G_2$, and a (full, partial, or empty) alignment $\mathcal{A}: V_1\rightarrow V_2$,
	find individual models $M_1$, $M_2$, common model $M_{12}$, 
	and transformations $\Delta_1$, $\Delta_2$ minimizing
	$L(M_{12}) + L(\Delta_1,\Delta_2) + L(G_1\parallel_{\mathcal{A}} G_2\mid M_{12},\Delta_1,\Delta_2)$.
}

\vspace*{0.7em}The search space is huge: 
Even if we searched for \emph{one} individual model only, 
limited the number of structures to $k$, 
set the minimum size of a structure to $r$, 
and required the union of all structures to form a partition of $V$,
we would need to search over $4^k$ times the number of partitions of $n$ into $k$ parts of size at least $r$.
These partitions are in bijection with the partitions of $n-k(r-1)$ into $k$ parts, 
and hence, there are $S(n-k(r-1),k)$ of them, where $S$ is the Stirling number of the second kind. 
Since we are looking for \emph{three} models with intricate interconnections, the search space for our problem is even larger---%
not to mention the NP-hard subproblems we need to solve to identify optimal structures (e.g., \textsc{MaxClique}).
Furthermore, our search space exhibits no structure such as (weak) (anti-)monotonicity of the total description length that would allow us to search it efficiently. 
Hence, we resort to heuristics.

\section{Algorithm}
\label{sec:algo}

We now introduce our second contribution, an algorithmic framework, called \ourmethod (\emph{Mo}del of \emph{mo}dels), to approximate the graph similarity description problem.
To discover good models in practice, we break this problem into two parts:
\begin{enumerate}
    \item Approximate the individual models $M_1$ and $M_2$ minimizing 
    $L(M_1) + L(G_1\mid M_1)~\text{and}~L(M_2) + L(G_2\mid M_2)$.
    Since these models can be thought of as graph summaries, 
    we refer to this task as \emph{graph summarization}.
    \item Given individual models $M_1$ and $M_2$, 
    approximate the common model $M_{12}$ and the associated transformations $\Delta_1$ and $\Delta_2$ minimizing
    $L(M_{12}) + L(\Delta_1,\Delta_2) + L(G_1\parallel_{\mathcal{A}} G_2 \mid M_{12}, \Delta_1, \Delta_2)$.
    Since we require there to be a unique structure in both $M_1$ and $M_2$ for each structure in $M_{12}$, 
    this means we search for an optimal alignment between the structures in $M_1$ and $M_2$. 
    Hence, we refer to this task as \emph{model alignment}.
\end{enumerate}
Given $M_1$, $M_2$, $M_{12}$, $\Delta_1$, and $\Delta_2$, the  NMD can be readily computed. 

Our architecture is flexible in that 
(1) any algorithm generating graph summaries using the structure vocabulary $\Omega$ can be used in the first step,
(2) any algorithm finding a common model and transformations based on individual graph summaries using the structure vocabulary $\Omega$ and the transformation vocabulary $\Sigma$ can be used in the second step, and
(3) all alphabets can be replaced with other alphabets (if they are mutually compatible and the encoding is suitably amended), just as the NMD can be substituted with an alternative measure, e.g.,  
to adapt our method to a specific domain.

\subsection{Step One: Graph Summarization (\oursummarizer)}
\label{subsec:find-summary}

We begin by summarizing each of our input graphs individually.
That is, our input is a single graph $G$ with node set $V$ and edge set $E$, 
and our output is a model $M$ approximately minimizing $L(M) + L(G \mid M)$.
Our procedure, called \oursummarizer, is given as Algorithm~\ref{alg:summarization}.

To start, we decompose our graph into a set $\mathcal{C}$ of connected components of diameter at most three (l.~\ref{line:decompose}).
We do this by iteratively selecting the node $v$ with the highest degree in the currently largest connected component to form a component $C\in \mathcal{C}$ with its neighbors, then deleting all edges incident with $v$, until no more components can be formed. 
This procedure is similar to the \textsc{SlashBurn} algorithm \cite{lim:14:slash}, but we recurse on the \emph{globally}, rather than the \emph{locally} largest connected component to ensure that all our components have small diameter.
The generated components are used as seeds to produce candidates for each structure type from our structure vocabulary $\Omega$,
where we merge candidates of the same type if they overlap on a large fraction of their nodes (l.~\ref{line:candidates-start}--\ref{line:candidates-end}).
We sort the remaining candidates, which can overlap on nodes \emph{and edges}, 
from largest to smallest (l.~\ref{line:sort}). 
Finally, for each structure $s$, in order, we add $s$ to $M$ if this reduces our description length (l.~\ref{line:add-start}--\ref{line:add-end}), i.e., if 
$L(s) + L(G\mid M \cup \{s\}) < L(G\mid M)$.

To generate a candidate of a certain structure type from a given component $C$ with node set $V_C$ (l.~\ref{line:candidates-generate}), we proceed as follows.

For a \emph{clique} with node set $V_s$, we first find the maximum clique in $C$ and include its nodes in $V_s$,
then we iteratively add the node from $V\setminus V_s$ with the highest degree in $G$ that is connected to at least $50\%$ of the nodes in $V_s$ until no more nodes fulfill this criterion.

For a \emph{star} with spoke set $V'_s$, 
we declare a node with the highest degree in $C$ to be the hub $v$, set $V'_s = V_C\setminus \{v\}$, 
and then iteratively (1) identify the nodes in $V'_s$ that have more than $0.05\cdot|V'_s|$ neighbors in $V'_s$,
and (2) remove the $\min\{(0.1+0.01i),1\}$ fraction of these nodes from $V'_s$ that has the most neighbors in $V'_s$ in iteration $i$.

For a \emph{biclique} with node sets $L$ and $R$, to start,
we set the right node set to be the (at most) $5$ nodes in a \emph{maximal} independent set (MIS) of $V_C$ that have the highest degree in $G$. 
We then identify the set $L' \subseteq V\setminus R$ of nodes that are connected to at least $50\%$ of the nodes in $R$, 
and set $L$ to be the (at most) $5$ nodes in an MIS of $L'$ that have the highest degree in $G$.
If $|L| < 3$ or $|R| < 5$, we discard the candidate early.
For the surviving candidates, we then iteratively 
(1) identify the set $X$ of nodes from $V\setminus (L\cup R)$ that are connected to at most $5\%$ of the nodes in $L$ and at least $50\%$ of the nodes in $R$, adding to $L$ the node from $X$ (if any) with the most neighbors in $R$, and
(2) perform (1), switching the roles of $L$ and $R$,
until no more nodes satisfy our criteria for addition to $L$ or $R$.

For a \emph{starclique} with node sets $L$ and $R$, to start,
we set $L$ to be the set of nodes contained in the maximum clique of $C$.
We then identify the set $R' \subseteq V\setminus L$ of nodes that are connected to at least $50\%$ of the nodes in $L$,
and set $R = \mathrm{MIS}(R')$.
Subsequently, we iteratively
(1) identify the set $X$ of nodes from $V\setminus (L\cup R)$ that are connected to at least $50\%$ of the nodes in $L$ and to at least $50\%$ of the nodes in $R$, adding to $L$ the node from $X$ (if any) with the most neighbors in $R$, and 
(2) identify the set $Y$ of nodes from $V\setminus (L\cup R)$ that are connected to at most $5\%$ of the nodes in $R$ and to at least $50\%$ of the nodes in $L$, adding to $R$ the node from $Y$ (if any) with the most neighbors in $L$,
until no more nodes can be added.

Running \oursummarizer on the graphs $G_1$ and $G_2$, 
we obtain interpretable individual models $M_1$ and $M_2$. 
Our next task is to align these models.

\subsection{Step Two: Model Alignment (\ouralignment)}
\label{subsec:find-alignment}

For the model alignment step, our inputs are the graphs $G_1$, $G_2$, 
the node alignment $\mathcal{A}$, 
and the models $M_1$, $M_2$. 
Our outputs are a common model $M_{12}$ and the transformations $\Delta_1$, $\Delta_2$, which together minimize $L(M_{12}) + L(\Delta_1,\Delta_2) + L(G_1\parallel_{\mathcal{A}} G_2~\mid~M_{12},\Delta_1,\Delta_2)$ approximately.
Our procedure, called \ouralignment, is given as Algorithm~\ref{alg:alignment}.

\begin{algorithm2e}[!t]
    \DontPrintSemicolon
\KwIn{Graph $G$; structure vocabulary $\Omega$}
\KwOut{Model $M$ with structure list $S$}
$\mathcal{C}\leftarrow$ Connected components of $G$ from decomposition\;\label{line:decompose}
$S', S\leftarrow [], []$\; 
\ForAll(\label{line:candidates-start}){\emph{structure types} $\omega\in \Omega$}{
    \ForAll{\emph{components} $C\in\mathcal{C}$}{
        Generate candidate of type $\omega$ from $C$\label{line:candidates-generate}
    }
    Merge generated candidates if they have large overlap\;
    Append remaining candidates to $S'$\;\label{line:candidates-end}
}
Sort structures $s \in S'$ by $(n_s,m_s)$ (descending)\;\label{line:sort}
\ForAll(\label{line:add-start}){\emph{structures} $s\in S'$}{
    \If{$L(s) + L(G\mid M\cup \{s\}) < L(G\mid M)$}{
        Append $s$ to $S$\label{line:add-end}
    }
}
\Return{$M$}
    \caption{Graph summarization with \oursummarizer}\label{alg:summarization}
\end{algorithm2e}

\begin{algorithm2e}[!t]
    \DontPrintSemicolon
    \KwIn{Individual models $M_1$, $M_2$ with structures $S_1$, $S_2$; node alignment~$\mathcal{A}$; transformation vocabulary $\Sigma$}
    \KwOut{Common model $M_{12}$ and transformations $\Delta_1$, $\Delta_2$ such that $\Delta_1(M_{12}) = M_1$, $\Delta_2(M_{12}) = M_2$}
    Compute constrained matching $\mathcal{M}\subseteq S_1 \times S_2$ \tcp*{Alg.~\ref{alg:matching}}\label{line:constrained-matching}
    $M_{12}, \Delta_1,\Delta_2 \leftarrow$ [], [], []\; 
    \ForAll(\label{line:common-start}){\emph{structures} $(s_1,s_2)\in\mathcal{M}$}{
        Compute the common structure $s$ for $(s_1,s_2)$\;
        Compute $\delta_i$ such that $\delta_i(s) = s_i$ for $i\in\{1,2\}$\;
        Append $s$ to $M_{12}$ and $\delta_i$ to $\Delta_i$ for $i\in\{1,2\}$\;\label{line:common-end}
    }
    \For(\label{line:leftover-start}){$i\in \{1,2\}$}{
        \ForAll{\emph{structures} $s\in S_i\setminus \{s\in S_i\mid \exists p \in\mathcal{M}: s\in p\}$}{
            Append $s$ to $\Delta_i$\label{line:leftover-end}
        }
    }
    \Return{$M_{12}$, $\Delta_1$, $\Delta_2$}
    \caption{Model alignment with \ouralignment}\label{alg:alignment}
\end{algorithm2e}

In the critical first step,
detailed below,
\ouralignment computes a (bipartite) matching $\mathcal{M}\subseteq S_1\times S_2$, pairing structures in $S_1$ with structures in $S_2$ (l.~\ref{line:constrained-matching}).
The matching is \emph{constrained} because we require that paired structures have the same type $\omega\in\Omega$. 
For each structure pair $(s_1,s_2) \in \mathcal{M}$, 
we then compute its common structure $s$ as well as transformations $\delta_1$ and $\delta_2$ such that $\delta_1(s) = s_1$ and $\delta_2(s) = s_2$, which we add to $M_{12}$, $\Delta_1$, and $\Delta_2$, respectively (l.~\ref{line:common-start}--\ref{line:common-end}).
Finally, we add the unpaired structures from both $S_1$ and $S_2$ to $\Delta_1$ and $\Delta_2$,
ensuring that $\Delta_1(M_{12}) = M_1$ and $\Delta_2(M_{12}) = M_2$ (l.~\ref{line:leftover-start}--\ref{line:leftover-end}).

Typically, the matching $\mathcal{M}$ is not uniquely defined. 
We are interested in the matching that helps us minimize the description length.
Sweeping the search space na\"ively is not an option: 
For a structure vocabulary $\Omega$, there exist
$\prod_{\omega\in\Omega}(\omega_{\max}-\omega_{\min})!\cdot\binom{\omega_{\max}}{\omega_{\min}}$
different \emph{maximal} matchings alone, 
where, for $f\in\{\min,\max\}$,
$\omega_f = f\{|\{s\in S_1 \mid \mathrm{type}(s) = \omega\}|,|\{s\in S_2 \mid \mathrm{type}(s) = \omega\}|\}$.
Hence, we propose a matching heuristic, \maximalgreedy, whose detailed pseudocode is given as Algorithm~3 in the Appendix.

If no node alignment is present, for $i\in \{1,2\}$, \maximalgreedy constructs \emph{node overlap graphs} $H_i$.
The nodes of these graphs are the structures in $S_i$, 
and the weights of their edges $F_i$ are the Jaccard similarities between the node sets of the structures (l.~3).
\maximalgreedy then builds a variant of the \emph{product graph} of $H_1$ and $H_2$, whose nodes are the subset of $S_1\times S_2$ that agrees on type, and whose edge weights are the product of the edge weights in $H_1$ and $H_2$ (l.~4--6).
\maximalgreedy then iteratively selects the heaviest edges in the product graph and removes all nodes that are incompatible with these edges (l.~7--12).
Finally, it pairs the remaining structures of the same type in descending order of their size (l.~13--19).

If a (partial) node alignment $\mathcal{A}$ is present, \maximalgreedy iteratively matches those structures $s_1$ and $s_2$ of the same type whose node sets have the largest average Jaccard similarity under $\mathcal{A}$ (l.~20--27).
For cliques, this equals the standard Jaccard similarity. 
For structures of other types, it is defined as
\begin{align*}
    \text{Jaccard}_{\mathcal{A}}(s_1,s_2) = \frac{1}{2}\cdot\sum_{i\in\{1,2\}}\frac{|\mathcal{A}(V_i(s_1))\cap V_i(s_2)|}{|\mathcal{A}(V_i(s_1))\cup V_i(s_2)|}\hphantom{\;},
\end{align*}
where $V_1$ and $V_2$ are the hub and spoke sets (for stars) or the left and right node sets (for bicliques and starcliques), respectively.

\maximalgreedy is designed to ensure interpretability:
In the presence of a node alignment, it honors the node overlap of structures \emph{between} graphs, 
and in the absence of such an alignment, it honors the node overlap of structures \emph{within} graphs, 
all while respecting the constraints imposed by the structure types.

\subsection{Computational Complexity}
\label{subsec:complexity}

Having specified \oursummarizer and \ouralignment as the main components of \ourmethod, we now analyze \ourmethod's complexity.
Here, we assume that the total number of structures is $\mathcal{O}(1)$, which is required for interpretability. 

For \oursummarizer, due to the set intersection operations involved, constructing structure candidates is $\tilde{\mathcal{O}}(nm)$, where $\tilde{\mathcal{O}}$ hides polylogarithmic factors.
To decide whether to add a candidate to our model, we need to find the maximum entropy distribution for the adjacency matrix of the graph given that model, which is $\mathcal{O}(1)$ since the number of Lagrange multipliers is $\mathcal{O}(1)$.
We also need to keep track of the mapping of Lagrange multipliers to potential edges, which is $\mathcal{O}(n^2)$ with $\mathcal{O}(1)$ candidates.
Hence, \oursummarizer runs in $\tilde{\mathcal{O}}(nm)$.

\ouralignment's complexity is driven by $\mathcal{O}(1)$ Jaccard similarity computations,
which together take $\mathcal{O}(n^2)$ in the worst case ($\mathcal{O}(n)$ on average), where $n = \max\{n_1,n_2\}$.
Given individual models $M_1$, $M_2$, and their model alignment $(M_{12},\Delta_1,\Delta_2)$, computing the NMD takes $\mathcal{O}(1)$ basic arithmetic operations, i.e., its total complexity is $\mathcal{O}(1)$.

Overall, \ourmethod's complexity is dominated by \oursummarizer, and hence $\tilde{\mathcal{O}}(nm)$ in the worst case.
However, as we show in Section~\ref{sec:exps}, in practice, \ourmethod's performance is near-linear in the number of edges.

\section{Related Work}\label{sec:related}

To the best of our knowledge, we are the first to treat graph similarity assessment primarily as a \emph{description} problem, 
rather than as a \emph{measurement} problem.
Related work broadly falls into two categories: 
graph similarity measurement and graph summarization.

\paragraph{Graph Similarity Measurement} 
Early work on graph similarity measurement uses \emph{global} measures that capture graph \emph{structure}, 
e.g., graph edit distance and maximum common subgraphs \cite{kaden:90:editdist,raymond:02:rascal,zeng:09:editdistapx}.
Later research also explores measures that capture graph \emph{connectivity} \cite{koutra:16:deltacon}, 
leverage graph \emph{decompositions} \cite{nikolentzos:2018:degeneracy}, 
or aggregate \emph{local} similarities via node feature distributions \cite{berlingerio:13:netsimile,bagrow:2019:portrait}.
Building on prior work concerning \emph{graph kernels} \cite{borgwardt:2005:path,shervashidze:09:kernels}, 
recent contributions investigate similarity learning via \emph{deep} graph kernels \cite{yanardag:15:deepkernels,togninalli:2019:wasserstein,ma:2019:survey,ok:2020:deep}.
 
In contrast to the existing literature,
first, our \emph{primary goal} is graph similarity \emph{description}, not \emph{measurement}.
Second, our \emph{perspective} emphasizes \emph{interpretability}, which leads us to build on intuitive meso-level structures, 
rather than (overwhelmingly numerous) micro-level node features, motifs, or (opaque) macro-level graph features.
Third, our \emph{approach} is novel in that it formalizes graph similarity as a model selection task using the MDL principle. 

When evaluating \ourmethod, we compare the NMD to another normalized similarity measure that is also based on information-theoret\-ic principles: 
the \emph{Network Portrait Divergence} (NPD) \cite{bagrow:2019:portrait}.
The NPD is the Jensen-Shannon divergence of the probability distributions of the input graphs that describe how many nodes have $x$ neighbors at distance $y$.
We show that the NMD and the NPD often capture similar trends, but only the NMD is intuitively interpretable.

\paragraph{MDL-Based Graph Summarization}
Although novel in graph similarity assessment, 
the MDL principle has been used extensively in graph summarization. 
Starting with the SUBDUE system \cite{cook:94:subdue}, 
a rich line of work has sought to move summarization beyond clustering using more expressive vocabularies to identify meaningful structures in \emph{static} graphs \cite{feng:13:cxprime,lim:14:slash,koutra:15:vog,goebl:16:megs}.
MDL has also been used to find partitions in graph \emph{streams} \cite{sun:07:graphevo} or structures ranging across multiple \emph{aligned} snapshots of \emph{dynamic} graphs \cite{shah:15:timecrunch,kapoor:2020:online}.

Going beyond the existing literature, 
first, we allow our structures to overlap not only on nodes but also on \emph{edges}, and we can handle multiple graphs even if they are \emph{unaligned}.
Second, we improve the \emph{methodology} of previous static summarization methods, leveraging more noise-tolerant structure definitions and an optimal encoding of the data under the model.
Third, in our structure search, we emphasize result \emph{quality}, 
reflecting the need for accurate graph summaries as inputs to our comparison algorithm.

When evaluating \ourmethod, we compare \oursummarizer to \textsc{VoG} \cite{koutra:15:vog}, a static graph summarizer built on a similar graph decomposition method and vocabulary of interpretable structures (including cliques, bicliques, stars, and chains) that neither uses maximum entropy modeling or component post-processing nor allows edge overlap.
We show that \oursummarizer discovers more informative summaries than \textsc{VoG}. 

\section{Experiments}
\label{sec:exps}

\newcommand{\qone}{Does \oursummarizer create useful graph summaries?}
\newcommand{\qtwo}{Does \ouralignment discover interpretable common models?}
\newcommand{\qthree}{Does \ourmethod yield informative similarity scores?}

We now present our third contribution, 
an extensive evaluation of the framework presented in Section~\ref{sec:algo}.
To this end, we implement \oursummarizer in Julia and all other parts of \ourmethod in Python. 
We run our experiments on Intel E5-2643 CPUs with 256~GB RAM.
All data, code, and results are publicly available.\!\footnote{%
\oururl; \ourdoi
}
We answer three questions:
\begin{enumerate}[label={\bfseries Q\arabic*}]
	\item \qone
	\item \qtwo
	\item \qthree
\end{enumerate}
To ensure interpretability, we limit our summaries to at most $100$ structures, although allowing more would give better compression.

In our experiments, we use real-world graphs from seven collections (cf. Appendix Table~3):
Graphs in the \emph{asb} and \emph{asp} collections represent peering relations between Autonomous Systems, each in $9$ different weeks from $2001$ \cite{leskovec:07:graph}; 
graphs in the \emph{bio} collection represent physical interactions between human proteins in $144$ different tissues, 
where the protein identities induce partial node alignments between all pairs of graphs in the collection \cite{zitnik:17:tissue};
graphs in the \emph{clg} and \emph{csi} collections represent arXiv collaboration networks of cs.LG and cs.SI in each year from $2011$ to $2020$ \cite{kaggle:2020:arxiv};
and graphs in the \emph{lus} and \emph{lde} collections represent references between sections of the United States Code and the Code of Federal Regulations or their German equivalents in each year from $1998$ to $2019$ \cite{coupette:2021:law}.
We also include two collections of synthetic random graphs, \emph{rba} and \emph{rer}, based on the Barab\'asi-Albert (BA) model and the Erd\H{o}s-R\'enyi (ER) model.
Our graphs vary in size and density, containing up to $160K$ nodes and up to $525K$ edges (cf. Appendix Figure~10).

\begin{figure}[!t]
	\centering
	\includegraphics[width=0.7\linewidth]{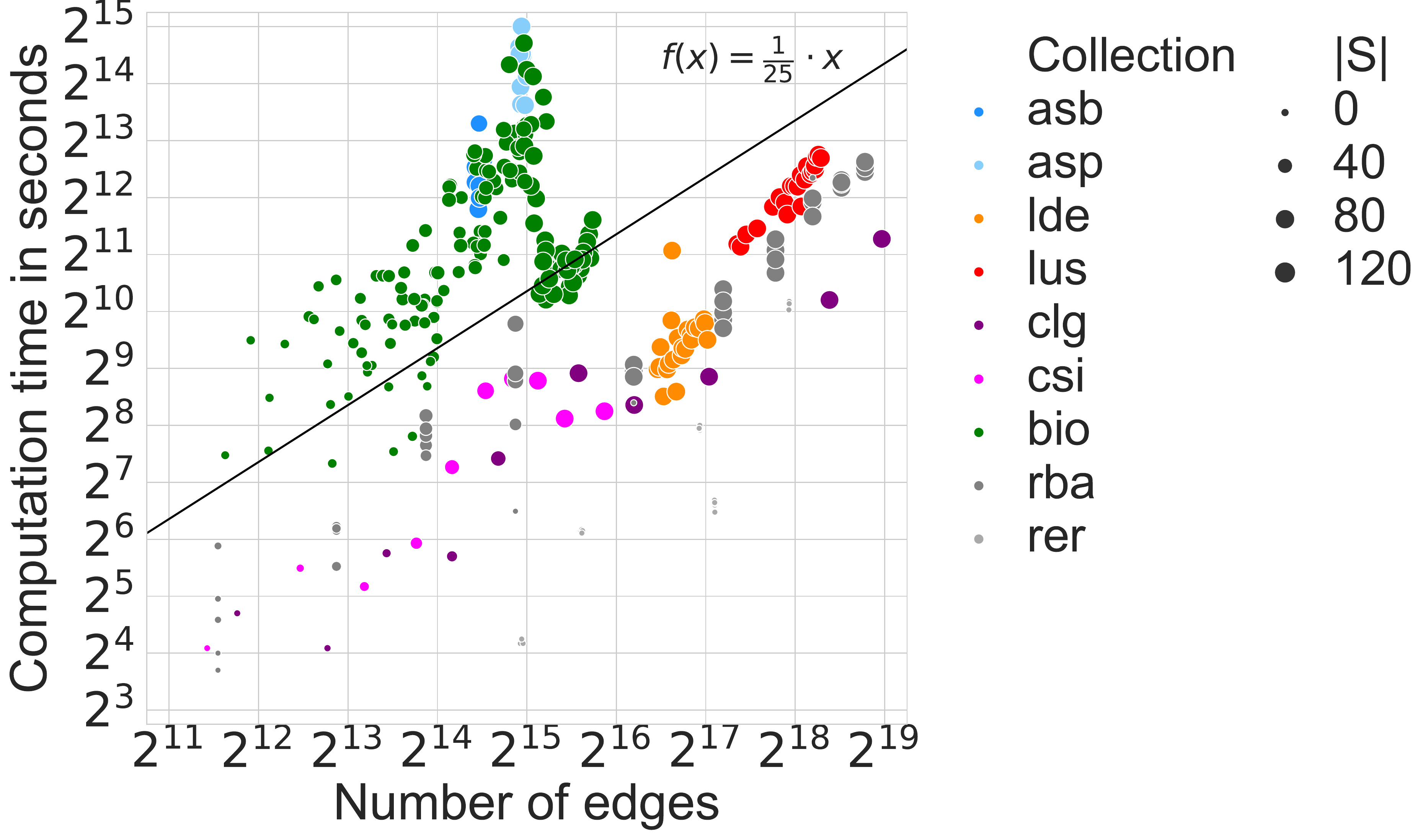}
	\vspace*{-9pt}\caption{%
	\oursummarizer is near-linear and output-sensitive.  
	Its computation time is shown as a function of $m$, with markers scaled by the number of discovered structures $|S|$.
	}\label{fig:performance}
\end{figure}

\paragraph{\bfseries\emph{Q1: \qone}}\label{q1}
In our context, graph summaries are useful if they capture the essence of a graph in an easily comprehensible manner.
To assess whether \oursummarizer creates such summaries, 
we start by comparing with \vog, which has been shown to produce useful graph summaries, 
on graphs from the \vog paper \cite{koutra:15:vog}. 
As shown in Table~\ref{tab:momo-vs-vog}, in all experiments, \oursummarizer saves more bits relative to the original encoding length than \vog-$k$ for the same $k$, i.e., it achieves a better compression $L\%$.
Moreover, \oursummarizer's compression is comparable to that of \vog-Greedy, 
although it uses much fewer structures.
That is, even though our encoding of the data under the model is optimal, 
we manage to save more bits per structure than \vog.
We also observe that while \oursummarizer uses its entire vocabulary to summarize its input graphs, \vog finds almost only stars.
As we show in Figure~\ref{fig:performance}, despite doing more work than \vog, 
\oursummarizer is near-linear in practice.

In the left panel of Figure~\ref{fig:structuretypes}, we tally how many structures of each type we find and what compression we achieve, on average, in each graph from our collections.
Since the edges of ER graphs are chosen uniformly at random, 
and BA graphs are grown using preferential attachment, 
it comes as no surprise that we find at most one star (with minimal gain) in ER graphs and only stars in BA graphs, achieving no or little compression.
The highest fraction of cliques occurs in the collaboration graphs (\emph{clg}, \emph{csi}), where papers with many authors induce cliques.
The hubs of the stars in these graphs correspond to well-known researchers with many independent collaborations, 
e.g., \emph{Yoshua Bengio}, \emph{Yang Liu}, and \emph{Sergey Levine} in \emph{clg} $2020$.
Some researchers occur in several structures, e.g., in \emph{csi} $2020$, $6$ of the spokes in the star around \emph{Christos Faloutsos}, shown in the right panel of Figure~\ref{fig:structuretypes}, 
reappear in the star around \emph{Danai Koutra}, 
and some of them are also connected.
This demonstrates the importance of allowing structures to overlap on nodes \emph{and} on edges, a feature absent from other state-of-the-art summarizers like \vog. 
For the law graphs (\emph{lde}, \emph{lus}), analysis by the first author (who happens to hold a PhD in law) and discussion with legal scholars
revealed that we can classify stars based on the ratio of the in- and out-degree of their hubs to uncover their legal function. 
Thus, \oursummarizer produces summaries that are useful to domain experts even for \emph{directed} graphs, which sets it further apart from other methods.

\begin{table}[!t]
	\centering
	\setlength{\tabcolsep}{3pt}
	\caption{%
	\oursummarizer compresses graphs more efficiently than \vog. $|S|$ is the number of structures, and $L\%$ is the compression (in percent of the uncompressed encoded length).
	}\label{tab:momo-vs-vog}
	\vspace*{-6pt}\begin{tabular}{lrrrrrrrr}
\toprule
 &&&\multicolumn{2}{c}{\bfseries\textsc{Beppo}}&\multicolumn{2}{c}{\bfseries \textsc{VoG}-$k$}&\multicolumn{2}{c}{\bfseries \textsc{VoG}-G}\\
 \cmidrule(lr){4-5}\cmidrule(lr){6-7}\cmidrule(lr){8-9}
 \textbf{Graph}   &   $n$ &    $m$ &   $|S|$ &   $L\%$ &   $|S|$ &   $L\%$ &   $|S|$ &   $L\%$ \\
\midrule
 Epinions                                                                                                                                                          & 75879 & 405740 &     100 &       20 &     100 &        5 &    2746 &       19 \\
 Enron                                                                                                                                                             & 79870 & 288364 &     100 &       18 &     100 &        7 &    2331 &       25 \\
 AS-Oregon                                                                                                                                                         & 13579 &  37448 &     100 &       28 &     100 &       21 &     399 &       29 \\
 Chocolate                                                                                                                                                         &  2877 &   5467 &      55 &        9 &     100 &        7 &     101 &       12 \\
 Controversy                                                                                                                                                       &  1093 &   2942 &      20 &       15 &     100 &        4 &      35 &       13 \\
\bottomrule
\end{tabular}
\end{table}

\addtolength{\belowcaptionskip}{-5pt}
\begin{figure}[!t]
	\centering
	\setlength{\tabcolsep}{3pt}
	\vspace*{-4pt}\begin{subtable}[b]{0.44\linewidth}
		\centering
		\begin{tabular}{lrrrrr}
\toprule
 \textbf{Coll.}   &   $\mathbf{\widehat{cl}}$ &   $\mathbf{\widehat{st}}$ &   $\mathbf{\widehat{bc}}$ &   $\mathbf{\widehat{sc}}$ &   $\widehat{L\%}$ \\
\midrule
 asb              &                         1 &                        96 &                         0 &                         1 &              29 \\
 asp              &                         3 &                        91 &                         0 &                         6 &              29 \\
 bio              &                         6 &                        52 &                         1 &                         2 &               9 \\
 clg              &                        24 &                        36 &                         0 &                         2 &               8 \\
 csi              &                        13 &                        48 &                         0 &                         1 &              16 \\
 lde              &                         0 &                        98 &                         1 &                         1 &               1 \\
 lus              &                         0 &                        95 &                         4 &                         1 &               4 \\
 rba              &                         0 &                        68 &                         0 &                         0 &               3 \\
 rer              &                         0 &                         1 &                         0 &                         0 &               0 \\
\bottomrule
\end{tabular}
		\subcaption*{Results of \oursummarizer}
	\end{subtable}%
	\begin{subfigure}[b]{0.56\linewidth}
		\centering
		\vspace*{-4pt}\includegraphics[width=\linewidth]{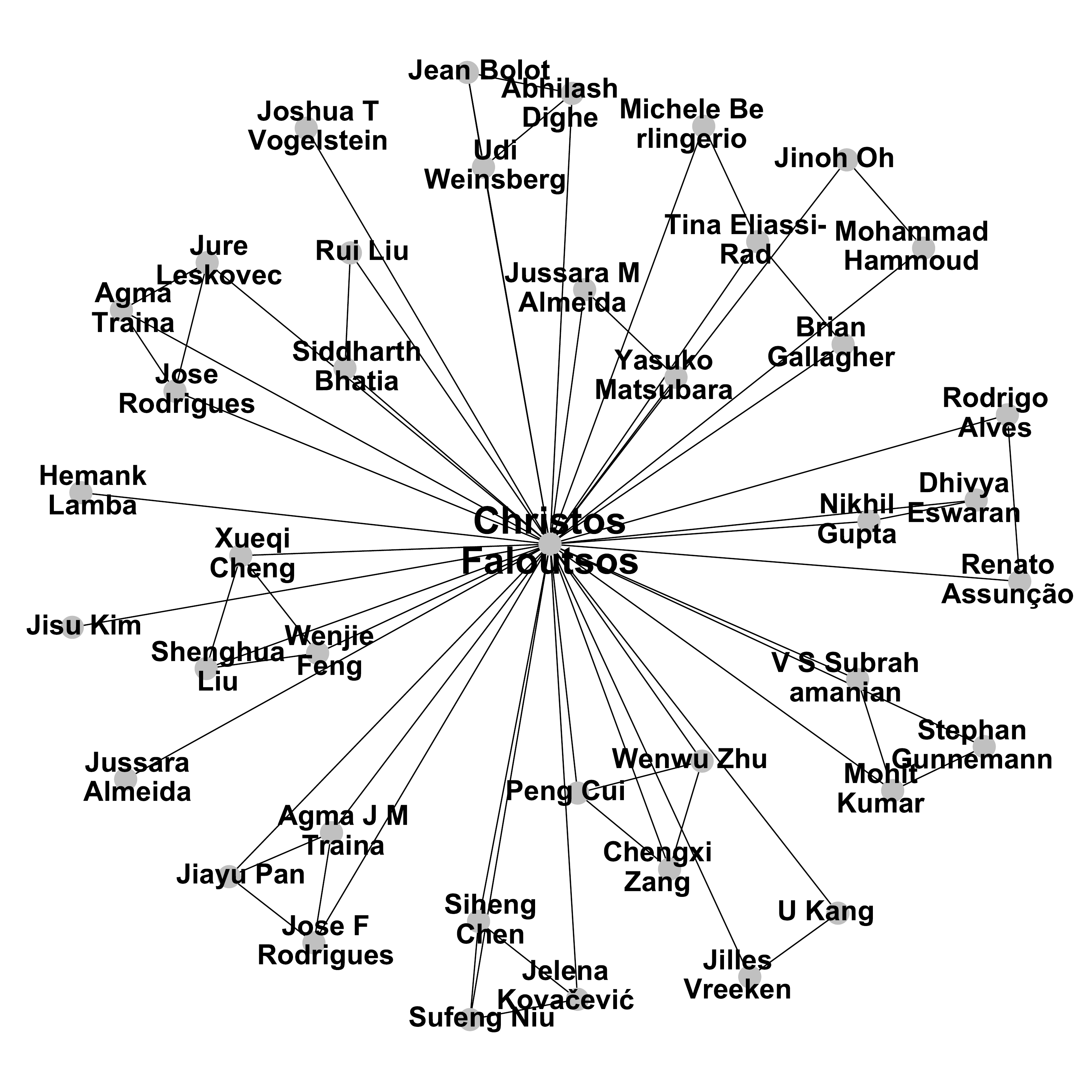}%
		\vspace*{-8pt}\subcaption*{Example star from \emph{csi} $2020$}
	\end{subfigure}%
	\vspace*{-6pt}\caption{%
	\oursummarizer identifies meaningful structures. 
	We show its average compression and number of structures per type (left), and an example star discovered in \emph{csi} $2020$ (right).}\label{fig:structuretypes}
\end{figure}
\addtolength{\belowcaptionskip}{5pt}

\begin{figure}[!t]
	\centering
	\includegraphics[width=\linewidth]{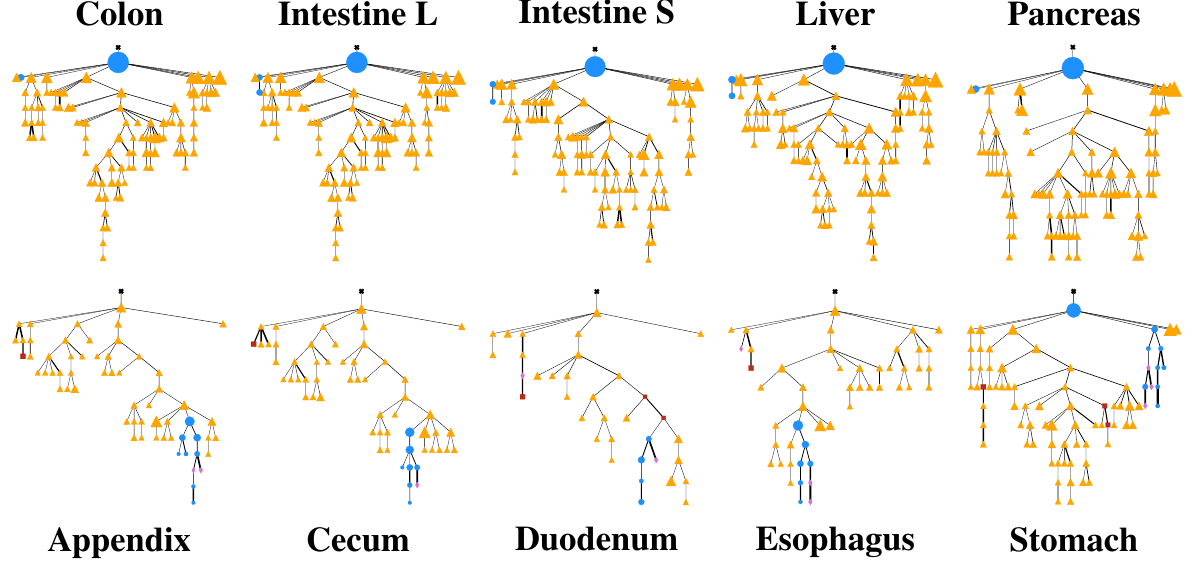}
	\vspace*{-18pt}\caption{%
	\oursummarizer creates similar summaries with similar node overlap structure for similar graphs.
	The node overlap trees for selected digestive tract organs in the \emph{bio} collection mirror the functional (dis)similarity between these organs.
	}\label{fig:bio-trees}
\end{figure}

Since we allow structures to overlap, \oursummarizer's summaries can be visualized intuitively as \emph{node overlap trees}.
Node overlap trees are the maximum spanning trees of node overlap graphs, i.e., each vertex in them represents a structure, the edge weights are the Jaccard similarities between the node sets of the structures, and we remove all edges that are lightest in a cycle. 
To ensure connectivity, we introduce a root vertex that connects to the vertex with the largest degree inside each component.
We depict the node overlap trees for selected digestive tract tissues from the \emph{bio} collection in Figure~\ref{fig:bio-trees}. 
Here, larger shapes indicate larger structures, and thicker edges indicate higher Jaccard similarities.
From the vertices and the connectivity structure of the trees, it is immediately apparent that the top-row tissues are very similar, 
and indeed, the functions performed by the organs they represent are closely related.

\begin{figure}[!t]
	\centering
	\includegraphics[width=\linewidth]{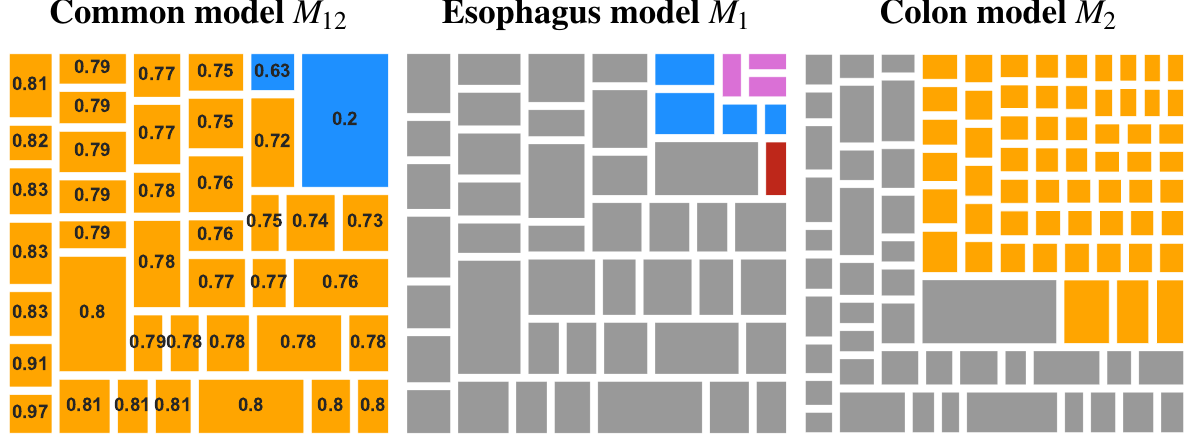}
	\vspace{-18pt}\caption{%
	\ouralignment discovers interpretable common models. 
	The common model (left) for esophagus (middle) and colon (right) contains mostly structures with high average Jaccard similarity (annotations).
	Each rectangle corresponds to a structure, sized proportionally to its number of nodes, and shared structures are greyed out in the individual models.
	}\label{fig:gigi-treemaps}
\end{figure}

\paragraph{\bfseries\emph{Q2: \qtwo}}\label{q2}
As \ouralignment builds on \oursummarizer, the common models it discovers are composed of easily comprehensible structures. 
By construction, this ensures a certain degree of interpretability.
To understand the composition of a common model $M_{12}$ and its relationship to individual models $M_1$ and $M_2$, we can further visualize these models using treemaps. 
We show an example from the \emph{bio} collection in Figure~\ref{fig:gigi-treemaps}, contrasting the individual models for esophagus and colon with their common model.
We see that esophagus and colon have many common structures, most of them stars, 
but the esophagus has more complex or dense structures (cliques, bicliques, and starcliques), while the colon has more simple sparse structures (stars).
Using the node alignments between the \emph{bio} graphs to annotate the shared structures with their average Jaccard similarities, we observe that all stars that are shared between esophagus and colon have a shared hub (indicated by a similarity above $0.5$).
Similar observations can be made for other tissues, e.g., the largest cliques in the top-row tissues from Figure~\ref{fig:bio-trees} all have a Jaccard similarity of at least $0.58$.
This indicates that \emph{housekeeping proteins} might be expressed as \emph{housekeeping structures} that recur across tissues,
but a detailed investigation of this hypothesis lies outside the scope of this paper.

\begin{figure}[!t]
	\centering
	\includegraphics[width=\linewidth]{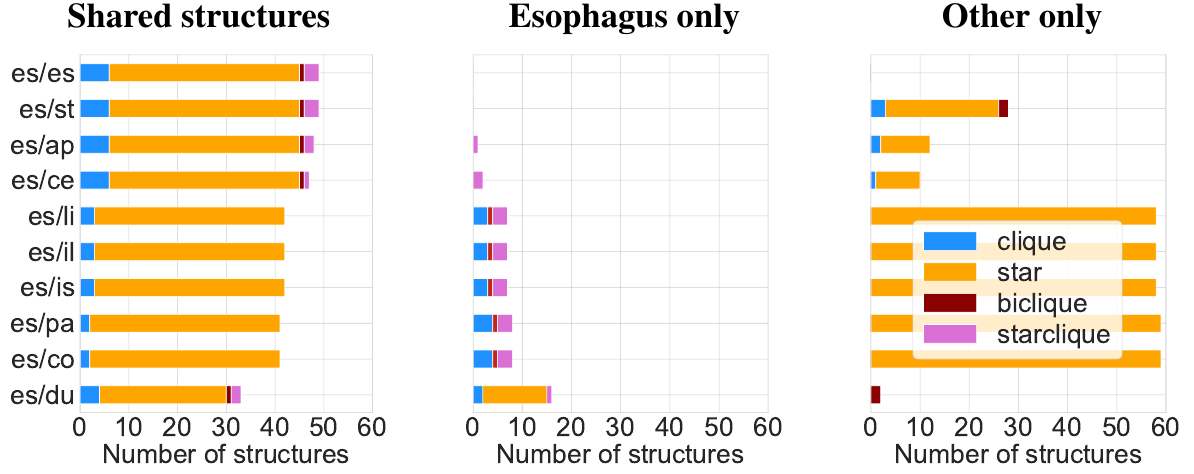}	
	\vspace*{-18pt}\caption{%
	\ouralignment allows multi-graph comparisons. 
	Here, we juxtapose shared (left) and specific (middle, right) structures for the esophagus and the tissues from Figure~\ref{fig:bio-trees}.
	}\label{fig:gigi-bio}
\end{figure}

Beyond bilateral graph similarity assessment, \ouralignment's output enables comparisons between multiple graphs. 
As an example, in Figure~\ref{fig:gigi-bio}, we display the composition of the common models for comparisons of the esophagus with the tissues from Figure~\ref{fig:bio-trees} as a triptych of stacked bar charts.
The graphic illustrates that the relationship between esophagus and colon, shown in Figure~\ref{fig:gigi-treemaps}, is comparable to that of the esophagus and \emph{any} top-row organ from Figure~\ref{fig:bio-trees},
and that all bottom-row organs share a biclique structure.

To further explore the relationships between shared structures, we can leverage \emph{common} node overlap graphs, i.e., node overlap graphs induced by our structure matching $\mathcal{M}$, with nodes $(s_1,s_2)\in \mathcal{M}$, edges $((s_1,s_2),(t_1,t_2))$, and edge weights $\prod_{i\in\{1,2\}}\mathrm{Jaccard}(s_i,t_i)$.
These graphs convey an interpretable notion of equivalence between the matched structures.
To visualize common node overlap graphs, we again use node overlap trees, 
and Figure~11 in the Appendix shows an example from the \emph{lde} collection.
While not all patterns from the individual trees recur in the common tree, the trees induced by the common tree in the individual node overlap graphs typically weigh a large fraction of the individual node overlap trees, i.e., the alignments discovered by \ouralignment respect much of the node overlap shared between the structures in our input graphs.

\paragraph{\bfseries\emph{Q3: \qthree}}\label{q3}
Although our focus is similarity \emph{description}, we can also use our similarity score, the NMD, for similarity \emph{measurement}.
As depicted in Figure~12 in the Appendix, experiments on synthetic models show that the NMD is almost scale-invariant when the graphs contain rescaled versions of the same structures and their size differs within one order of magnitude, with larger size differences leading to larger NMD values.
The NMD also behaves intuitively for models of varied compositions, showing a strong correlation with the number of structures that can be matched across graphs.

When we compare NMDs to \emph{Network Portrait Divergence} values (NPDs), on the yearly snapshots of the IBM GitHub collaboration network from $2013$ to $2017$ used in \cite{bagrow:2019:portrait},
the general trends are quite similar,
but some years are \emph{more} similar and others are \emph{less} similar under NMD than under NPD (cf. Appendix Figure~13).
However, only our results are also interpretable: 
In $2014$, for example, the network only has one star structure, 
explaining its high dissimilarity to $2015$, which features one starclique and two cliques. 
The differences between NMD values and NPD values are likely due to the dependence of NPD on graph size, but since the underlying statistics are not intuitively comprehensible, we cannot be sure.

\begin{figure}
	\centering
	\includegraphics[height=2.4cm]{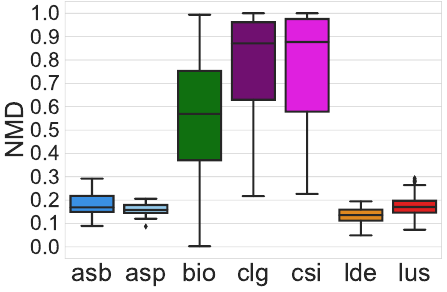}
	\vspace*{-9pt}\caption{%
	NMDs are lower for cross-temporal comparisons of systems experiencing gradual change (\emph{asb}, \emph{asp}, \emph{lde}, \emph{lus}) than for cross-temporal comparisons of systems undergoing radical change (\emph{clg}, \emph{csi}) or cross-sectional comparisons (\emph{bio}).
	}\label{fig:nmd-summary-statistics}
\end{figure}

\begin{figure}
	\includegraphics[height=2.4cm]{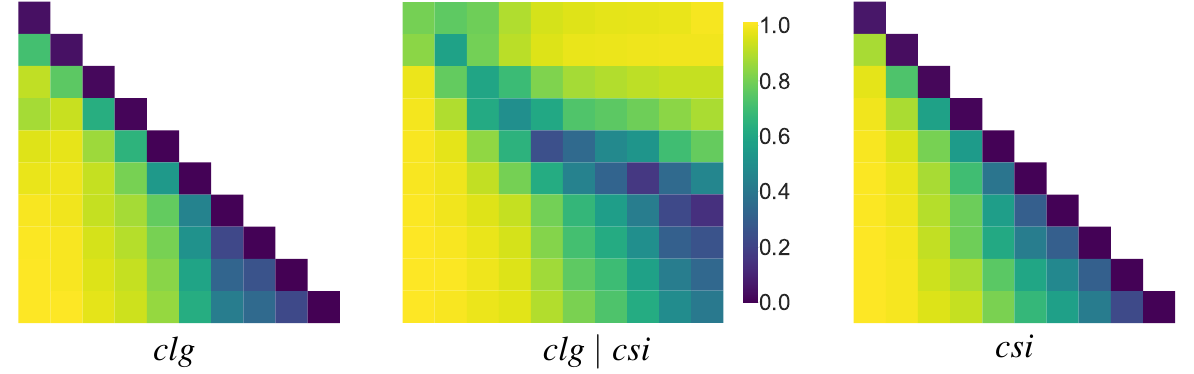}
	\vspace*{-9pt}\caption{
	NMDs yield nuanced insights.
	The NMDs for the \emph{clg} and \emph{csi} graphs from $2011$ (top/left) to $2020$ (bottom/right) show the arrow of time within each collection (left, right) and the lag between \emph{clg} and \emph{csi} from $2015$ onwards (middle).
	}\label{fig:nmd-heatmaps}
\end{figure}

In Figure~\ref{fig:nmd-summary-statistics}, we depict the distribution of NMDs for all pairwise comparisons of \emph{different} graphs in our real-world collections. 
We see that NMDs span the whole range, and their distribution differs depending on the type of comparison (\emph{cross-sectional} vs. \emph{cross-temporal}) and the type of change (\emph{gradual} vs. \emph{radical}) experienced by the system we study.
To illustrate radical change, we show the NMDs of the collaboration graphs (\emph{clg}, \emph{csi}) from $2011$ to $2020$ in Figure~\ref{fig:nmd-heatmaps}.
Both collections display the arrow of time, but self-similarity drops faster in \emph{clg} than in \emph{csi} from about $2015$ onwards,
and when comparing across collections, \emph{csi} $2015$ is most similar to \emph{clg} $2015$ but \emph{csi} $2020$ is most similar to \emph{clg} $2017$.
Thus, while both communities have picked up tremendous pace in the past ten years, development in \emph{clg} has been measurably more rapid than in \emph{csi}.

\section{Conclusion}\label{sec:conclusion}

We study graph similarity assessment as a \emph{description} problem, 
guided by the question ``\emph{how} are these graphs similar?''.
Formalizing the problem using the MDL principle, 
we capture the similarity of the input graphs in their \emph{common model} and the differences between them in \emph{transformations} to \emph{individual models}.
Since our search space is huge and unstructured, we propose a framework, \ourmethod, which breaks the problem into two parts:
\oursummarizer creates graph summaries that are useful to domain experts, and \ouralignment discovers interpretable common models, 
from which we can also derive informative similarity scores.
Through experiments on undirected and directed graphs of radically varying sizes from diverse domains, we confirm that \ourmethod works well and is near-linear in practice.

However, \ourmethod also leaves room for improvement.
For example, we would like to handle richer graph types, including weighted and attributed graphs, using encodings that fully leverage the available information.
Ideally, \oursummarizer and \ouralignment would discover their structure and transformation vocabularies on the fly, integrating domain-specific background knowledge in the process.
An improved structure encoding might account for the overlap between structures, which is currently considered explicitly only by \ouralignment.
Our NMD score focuses on the models of the input graphs, 
and a more comprehensive measure could integrate the data under these models.

Finally, MDL forces us to take a binary decision when considering structure candidates, which can result in large differences between models based on small differences between description lengths. 
To eliminate these artifacts and still retain interpretability, we could consider the full set of high-quality structure candidates and compress it using \emph{structures of structures}. 
This could lead to an \emph{interpretable graph kernel}, which---like overcoming \ourmethod's other limitations---%
constitutes an engaging topic for future work.


\bibliographystyle{ACM-Reference-Format-initialsonly}
\bibliography{bib/abbrev,bib/bib-jilles,bib/bib-paper}

\clearpage
\appendix
\section{Appendix}
\label{sec:apx}

In this Appendix, we provide further details on our algorithms, our data (i.e., graph collections), and our experiments. 
The basic notation used throughout our paper is summarized in Table~\ref{tab:symbols}.
We make all our data, code, and results publicly available.\!\footnote{\oururl; \ourdoi}

\subsubsection*{Algorithms}

In the following, we provide implementation details for all components of \ourmethod: \oursummarizer, \ouralignment, and the NMD computation.

\paragraph{\oursummarizer}
\oursummarizer has a size threshold, which allows us to stop decomposing connected components or discard generated candidates when they are too small. 
We set this threshold to $10$ for all our experiments except when comparing NMDs with NPDs, where we set it to $3$ because the input graphs are relatively small.

When deciding whether to merge candidates due to large overlap between their node sets in the final candidate generation step, 
we choose our merge thresholds such that we can reduce redundancy amongst candidates without harming structure quality.
For cliques, we set the merge threshold to $90\%$ of the nodes. 
For bicliques and starcliques, we require both the left sets and the right sets of two candidates to overlap on $90\%$ of the nodes.
We do not merge stars even for large overlaps because this would result in structures of a different type, which we generate separately.

We allow \oursummarizer to stop early if (1) it has added a given maximum number of structures to our model, 
or (2) we have tested a given maximum number of candidates \emph{without} adding them to our model. 
As described at the beginning of Section~6, to guarantee that our summaries are interpretable, we set their maximum number of structures to $100$. 
We set the maximum number of rejected candidates to $300$, but in our experiments, this becomes relevant only for graphs from the \emph{bio} collection. 
Because these graphs are relatively dense, \oursummarizer creates many overlapping candidates, but few of them suffice to cover most of the nodes and edges.
With early stopping, we can thus shorten the running time of \oursummarizer without compromising the quality of our graph summaries.

\paragraph{\ouralignment}
In Section~4.2, we give a verbal description of the \maximalgreedy matching heuristic used by \ouralignment.
Supplementing this description, we provide the detailed pseudocode of \maximalgreedy as Algorithm~\ref{alg:matching}.
To speed up the computation when no node alignment is given and structures do not overlap, 
our implementation has a no-overlap flag which, when set, allows us to skip directly to the greedy matching (l.~\ref{line:matching-greedy-start}--\ref{line:matching-greedy-end}).

\paragraph{NMD Computation}
If we compute the NMD na\"ively, it is in rare cases possible to obtain a value above $1$.
This occurs when the models for the two graphs are so different that encoding them individually is cheaper than encoding them using a common model and transformations, i.e., when $L(M_{12}) + L(\Delta_1,\Delta_2) > L(M_1) + L(M_2)$. 
As any value above $1$ signals that we do not gain any bits by compressing $G_1$ and $G_2$ together, we set the NMD to $1$ in this situation. 

For the \emph{bio} collection, the NMD distribution we report in Figure~8 is based on structure matchings using node alignments induced by protein identities. 
For all other collections, the distributions reported are based on structure matchings without node alignments.

\begin{table}[!t]
	\centering
	\caption{Basic notation.}\label{tab:symbols}
	\vspace*{-6pt}\begin{tabular}{p{0.2\columnwidth} p{0.7\columnwidth}}
    \toprule
    \bfseries Symbol&\bfseries Description\\
    \midrule
    $G_i = (V_i,E_i)$ & graph $i$ with node set (edge set) $V_i$ ($E_i$)\\ 
    $n_i = |V_i|$ & number of nodes in $G_i$\\
    $m_i = |E_i|$ & number of edges in $G_i$\\
    $\mathbf{A}_i$ & adjacency matrix of $G_i$\\
    $\mathcal{A}_{ij}$&alignment between $V_i$ and $V_j$\\
    \midrule
    $L(x)$ & number of bits to describe $x$ using our encoding\\
    $L_{\mathbb{N}}(x)$ & number of bits to describe $x$ using the universal code for integers\\
    $\log$&binary logarithm with $\log(0) = 0$\\
    $\lfloor x \rceil$&$x$ rounded to the closest integer\\
    \bottomrule
\end{tabular}
\end{table}

\begin{algorithm2e}[!t]
    \DontPrintSemicolon
    \SetKw{Break}{break}
    \KwIn{Structure lists $S_1$, $S_2$; node alignment $\mathcal{A}$}
    \KwOut{Structure matching $\mathcal{M}\subseteq S_1\times S_2$}
    
        $\mathcal{M} \leftarrow \emptyset$\; 
        \eIf(){$\mathcal{A} = \emptyset$}{
            $H_i \leftarrow (S_i,F_i,w_i)$ for $i\in\{1,2\}, w_i((s,t)) = \mathrm{Jaccard}(s,t)$\;\label{line:node-overlap}
            $V \leftarrow \big\{(s_1,s_2)\in S_1\times S_2 \mid \mathrm{type}(s_1) = \mathrm{type}(s_2) \big\}$\;\label{line:product-graph-start}
            $E \leftarrow \big\{\big((s_1,s_2),(t_1,t_2)\big) \mid (s_1, t_1)\in F_1, (s_2,t_2) \in F_2\big\}$\;
            $G$~\mbox{$\leftarrow (V,E,w), w\big(\big((s_1,s_2),(t_1,t_2)\big)\big) = \prod_{i\in\{1,2\}}w_i((s_i,t_i))$}\;\label{line:product-graph-end}
            \vspace*{-12pt}\While(\label{line:select-product-graph-edges-start}){$E\neq \emptyset$}{
                $(u,v) \leftarrow \arg\max_{(u,v)\in E} w((u,v))$\;
                Add $u$ and $v$ to $\mathcal{M}$\;
                $X \leftarrow \{x \in V \setminus \mathcal{M} \mid (x \cap u \neq \emptyset) \vee (x \cap v \neq \emptyset)\}$\;
                $E \leftarrow E \setminus \{(u,v)\}$\; 
                $G \leftarrow G[V\setminus X]$\;\label{line:select-product-graph-edges-end}
            }
            $\overline{S}_i \leftarrow S_i\setminus \{s \in S_i \mid \exists p\in \mathcal{M}: s\in p\}$ for $i\in\{1,2\}$\;\label{line:matching-greedy-start}
            \ForAll(){\emph{structures} $s_1\in \overline{S}_1$}{
            \ForAll(){\emph{structures} $s_2\in \overline{S}_2$}{
                \If(){$\mathrm{type}(s_1) = \mathrm{type}(s_2)$}{
                    Add $(s_1,s_2)$ to $\mathcal{M}$\;
                    $\overline{S}_i \leftarrow \overline{S}_i \setminus \{s_i\}$ for $i\in\{1,2\}$\;
                    \Break\;\label{line:matching-greedy-end}
                }
            }
        }
        }{\label{line:matching-align-start}
            $\overline{S}_i \leftarrow S_i$ for $i\in\{1,2\}$\; 
            \While(){%
                \textbf{\emph{true}}
            }{
                $U \leftarrow \{(s_1,s_2) \in \overline{S}_1 \times \overline{S}_2 \mid \mathrm{type}(s_1) = \mathrm{type}(s_2)\}$\;
                \lIf{$U = \emptyset$}{%
                    \Break
                }
                $(s_1,s_2) \leftarrow \arg\max_{(s_1,s_2)\in U}~\text{Jaccard}_{\mathcal{A}}(s_1,s_2)$\;
                Add $(s_1,s_2)$ to $\mathcal{M}$\; 
                $\overline{S}_i \leftarrow \overline{S}_i\setminus \{s_i\}$ for $i\in \{1,2\}$\;\label{line:matching-align-end}
            }
             
        }
        
        \Return{$\mathcal{M}$}
    \caption{Structure matching with \maximalgreedy}\label{alg:matching}
\end{algorithm2e}

\clearpage

\subsubsection*{Data}

\begin{figure}[t]
	\centering
	\begin{subfigure}{0.49\linewidth}
		\centering
		\includegraphics[width=\linewidth]{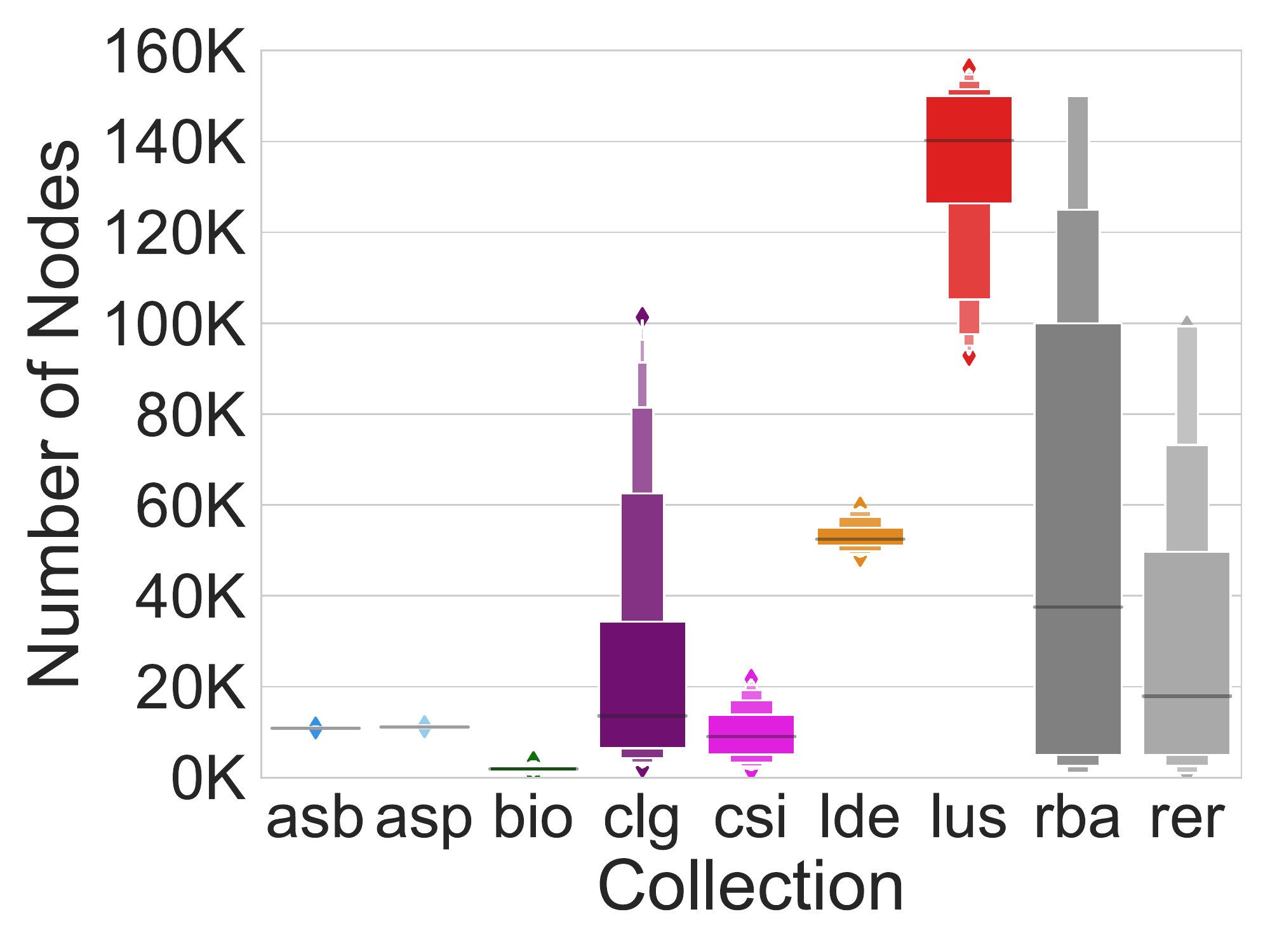}
		\vspace*{-14pt}\subcaption*{Distribution of $n$}
    \end{subfigure}~%
	\begin{subfigure}{0.49\linewidth}
		\centering
		\includegraphics[width=\linewidth]{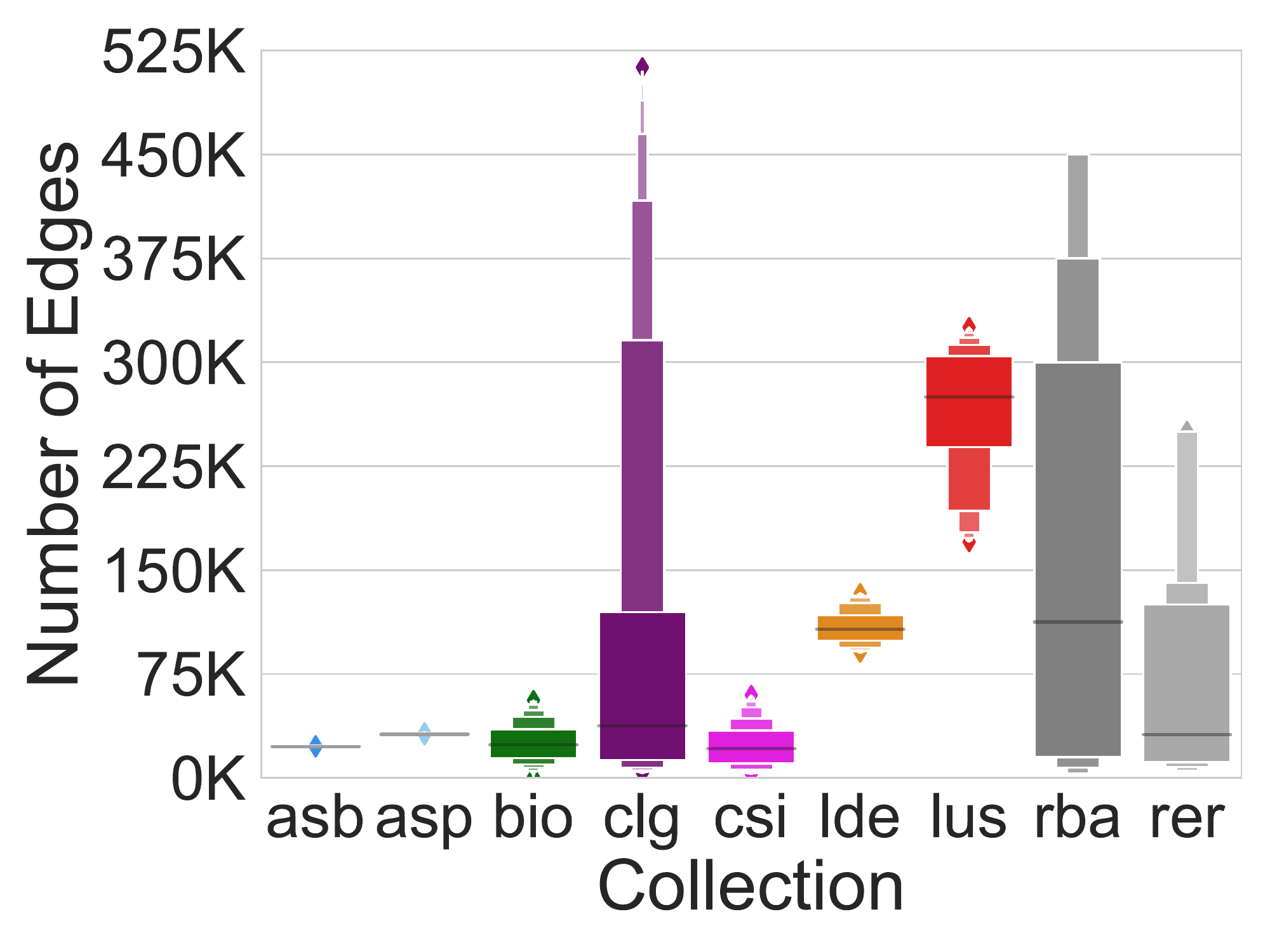}
		\vspace*{-14pt}\subcaption*{Distribution of $m$}
    \end{subfigure}
    \vspace*{-6pt}\caption{We consider graphs of radically varying sizes.}\label{fig:datasets}
\end{figure}

Beyond the implementation details of our algorithms, to facilitate the interpretation of our results, we provide additional background on the graph collections we use in our experiments. 
Supplementing the description at the beginning of Section~6, we give an overview of our graph collections in Table~\ref{tab:datasets}, 
and show their distributions of $n$ and $m$ in Figure~\ref{fig:datasets}.
For all collections except \emph{asb} and \emph{asp}, we perform some preprocessing to transform the data provided into the graphs we use, which is documented in our codebase. 
All random graphs are generated with graph generators available in the Python library \texttt{networkx}.

\begin{table}[t]
	\centering
	\setlength{\tabcolsep}{3pt}
	\small
	\caption{%
	Our experiments are based on graph collections from highly diverse domains. 
	$N$ is the number of networks in the respective collection.}\label{tab:datasets}
	\vspace*{-6pt}	\begin{tabular}{llrlc}
		\toprule
		\bfseries Coll.&\bfseries Description&\bfseries $N$&\bfseries Distinction&\bfseries Source\\
		\midrule
		asb &AS Oregon RouteViews basic&$9$&2001/03/31--05/26,&\multirow{2}{*}{\cite{leskovec:07:graph}}\\
		asp  &AS Oregon RouteViews plus&$9$&weekly&\\
		\midrule
		bio&physical protein interactions&144&human tissues&\cite{zitnik:17:tissue}\\
		\midrule
		clg &arXiv cs.LG collaborations&$10$&2011--2020,&\multirow{2}{*}{\cite{kaggle:2020:arxiv}}\\
		csi &arXiv cs.SI collaborations&$10$&yearly (11/01)&\\
		\midrule
		lde &German federal law&$22$&1998--2019,&\multirow{2}{*}{\cite{coupette:2021:law}}\\
		lus &United States federal law&$22$&yearly&\\
		\midrule
		rba&Barab\'asi-Albert random graphs&$50$&$10$ sizes,&--\\
		rer&Erd\H{o}s-R\'enyi random graphs&$50$&$5$ seeds&--\\
		\bottomrule
	\end{tabular}

\end{table} 

\subsubsection*{Experiments}

We complete our additional remarks by delivering the details we deferred in Section~6.
Full results for all our collections, including further visualizations, are provided along with our code.

When comparing \oursummarizer with \vog in Table~1 of Q1, we state the $n$ and $m$ we found in the original input data, which sometimes differs slightly from those reported in \cite{koutra:15:vog}.

\begin{figure}[!t]
	\centering
	\includegraphics[width=\linewidth]{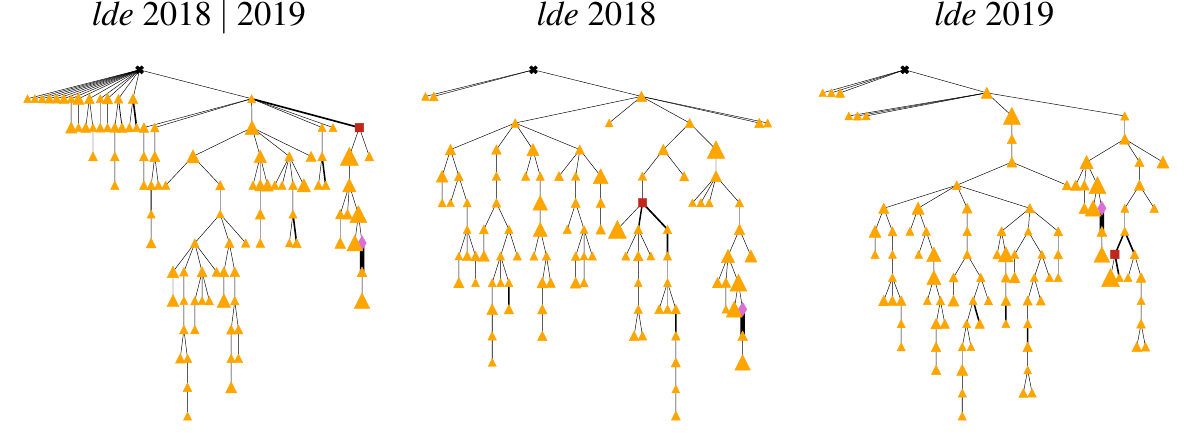}	
	\vspace*{-14pt}\caption{%
	\ouralignment discovers common models retaining much of the node overlap shared by the structures in the individual graphs, 
	as can be seen by comparing common (left) and individual (middle, right) node overlap trees.
	}\label{fig:gigi-law}
\end{figure}

As promised when answering Q2, we juxtapose common and individual node overlap trees for an example from the \emph{lde} collection in Figure~\ref{fig:gigi-law}. 
Here, the trees induced by the common tree in the individual node overlap graphs weigh more than $4/5$ of the individual node overlap trees.

\begin{figure}[!t]
	\includegraphics[width=\linewidth]{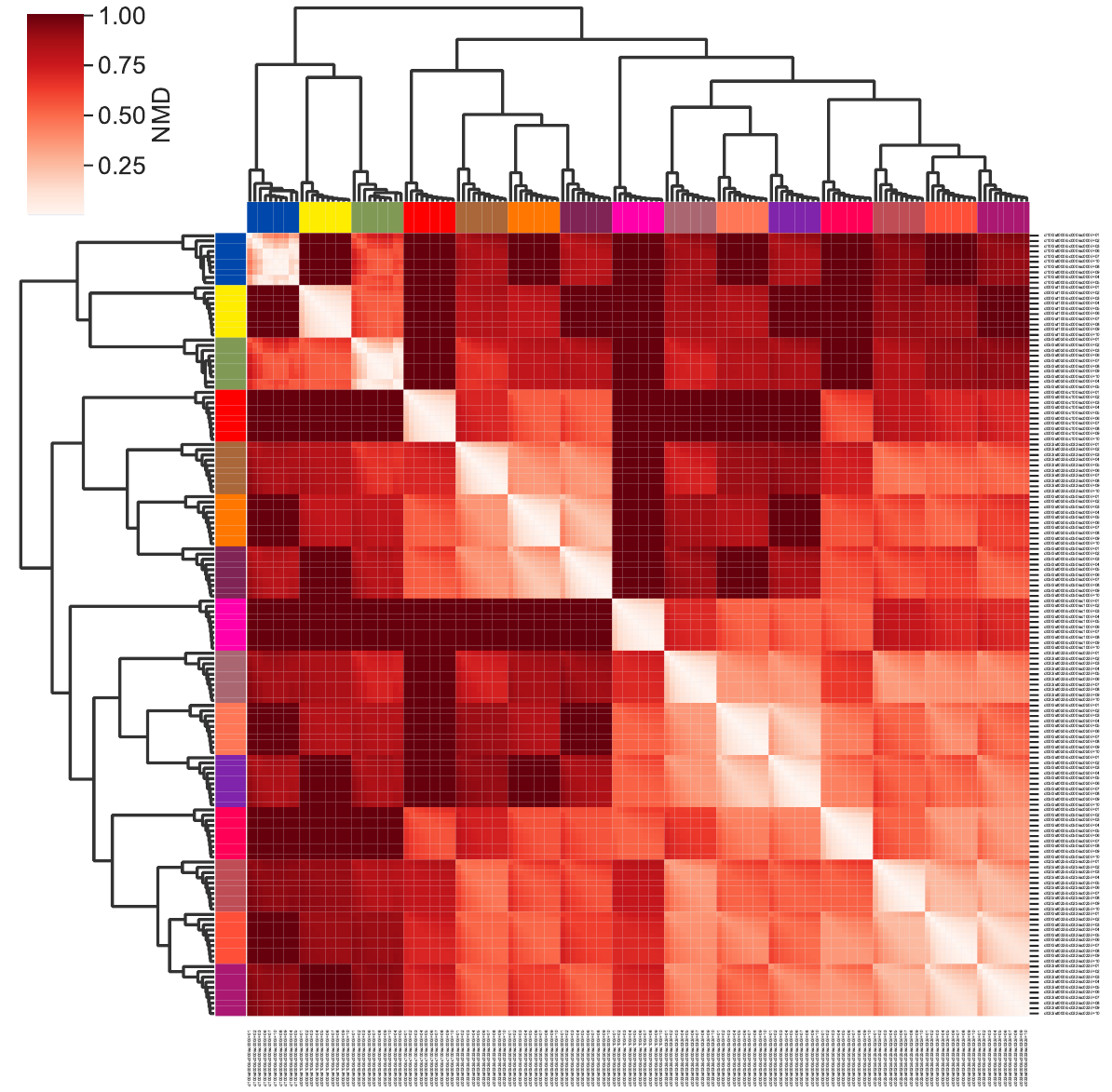}
	\vspace*{-14pt}\caption{%
	NMDs are (almost) scale-invariant (light strip along the diagonal) and correlate strongly with the number of structures that are matched across graphs (seven distinct shades of red).
	Row and column colors indicate model composition (mixed proportionally using blue, yellow, red, and magenta as the base colors for our structures); 
	labels show structure counts per type and graph size (represented by $i$).
	}\label{fig:nmd-synthetic}
\end{figure}

\begin{figure}[!t]
	\centering
	\includegraphics[width=\linewidth]{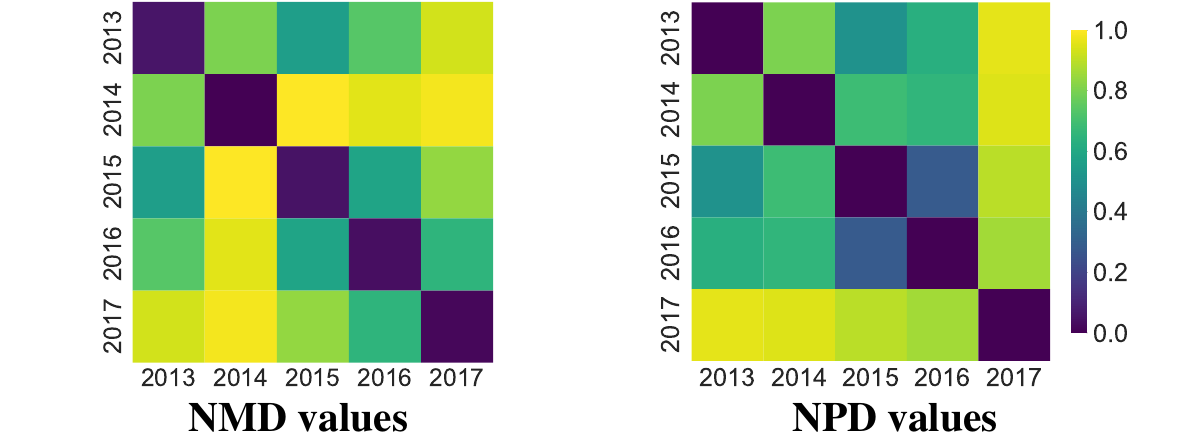}
	\vspace*{-14pt}\caption{%
	NMD and NPD detect similar trends, but where they differ, only NMD values are easy to interpret. 
	Here, we compare NMD values (left) with NPD values (right) on the IBM GitHub collaboration network from \cite{bagrow:2019:portrait}.
	}\label{fig:nmd-vs-npd}
\end{figure}

Supplementing the discussion in Q3, in Figure~\ref{fig:nmd-synthetic}, we provide a single-linkage hierarchical clustering of the NMDs of synthetic graphs with $n\in \bigcup_{i=1}^{10}\{i \cdot 10^4\}$ nodes
that contain $\lfloor 100/|\mathcal{S}|\rfloor$ structures of each type in $\mathcal{S}$, for $\mathcal{S} \in \mathcal{P}(\Omega) \setminus \emptyset$
($150$ graphs in total).
Finally, we visualize our comparison of NMDs with NPDs in Figure~\ref{fig:nmd-vs-npd}.

\clearpage

\end{document}